\documentclass[symmetry,article,submit,moreauthors,pdftex]{Definitions/mdpi}

\firstpage{1} 
\pubvolume{xx}
\issuenum{1}
\articlenumber{5}
\pubyear{2020}
\copyrightyear{2020}
\history{Received: date; Accepted: date; Published: date}

\usepackage{color}

\Title{Coupling hadron-hadron thresholds within a chiral quark model approach}

\Author{Pablo G. Ortega $^{1,\dagger}$\orcidA{} and David R. Entem $^{2,\dagger}$\orcidB{}}

\address{ 
$^{1}$ \quad Dpt. Fundamental Physics and IUFFyM, U. Salamanca, E-37008 Salamanca, Spain; pgortega@usal.es \\
$^{2}$ \quad Grupo de F\'\i sica Nuclear and IUFFyM, U. Salamanca, E-37008 Salamanca, Spain; entem@usal.es}

\corres{Correspondence: entem@usal.es}

\firstnote{These authors contributed equally to this work.}

\abstract{Heavy hadron spectroscopy was well understood within the naive quark model until
the end of the past century. However, in 2003, the $X(3872)$ was discovered, with puzzling
properties difficult to understand in the simple naive quark model picture. This state made
clear that excited states of heavy mesons should be coupled to two-meson states in order to
understand not only the masses but, in some cases, unexpected decay properties. In this work
we will review how the naive quark model can be complemented with the coupling to two hadron
thresholds. This program has been already applied to the heavy meson spectrum with the chiral
quark model and we show some examples where thresholds are of special relevance.
}


\begin{document}

\section{Introduction}

Heavy hadron spectroscopy started in November 1974, when
Brookhaven National Laboratory announced the discovery of a new
particle called $J$~\cite{PhysRevLett.33.1404} and, at the same time, the Stanford Linear Accelerator
reported the existence of another new particle, called $\psi$~\cite{PhysRevLett.33.1406}. 
Very soon both particles were seen as the same state, which we know now as the $J/\psi$ state of the
charmonium spectrum. This state was understood as a $q\bar q$ bound state in the naive quark model
and its discovery was the confirmation of the existence of the charm quark, that was predicted
by the GIM mechanism~\cite{PhysRevD.2.1285} few years before.

Afterwards, heavy meson spectroscopy developed very fast during the following years. The $\Upsilon(1S)$, the
first state with bottom quarks discovered, was found at Fermilab~\cite{PhysRevLett.39.252} in 1977.
Already at 1980 there were 11 new mesons included in the Particle Data Group table~\cite{RevModPhys.52.S1}
on these energy ranges, but that rate decreased to only 15 new states added during the period 1980 to 2003~\cite{PhysRevD.66.010001}.
However, during these last 17 years 35 new states have been added~\cite{10.1093/ptep/ptaa104},
considering only unflavored mesons.

In the case of the baryon spectra, the first evidences of charmed baryons came six months after the discovery of the
$J/\psi$, in 1975~\cite{PhysRevLett.34.1125}, but in 1980 only the $\Lambda_c$ and $\Sigma_c$ baryons where included
in the heavy baryon spectrum of the PDG~\cite{RevModPhys.52.S1}. In 2003, only 14 states were 
identified~\cite{PhysRevD.66.010001}, while again during the last 17 years 33 new states have been 
included~\cite{10.1093/ptep/ptaa104}. 

This impressive development of the heavy hadron spectrum has been possible thanks in part to the so 
called $B$-factories, like Belle and BaBar, which are electron-positron colliders tuned to the center 
of mass energy of the $\Upsilon(4S)$ that decays into two $B$ mesons. Other facilities like BESIII, with a lower energy
electron-positron collider, have contributed. Many impressive results have also been obtained and are underway
at LHC by the LHCb, CMS and ATLAS Collaborations. The next generation Super$B$-factory Belle II is 
running from 2018, and it is expected to give many important contribution to heavy hadron physics.

From a theoretical point of view, already in 1978, the Cornell model~\cite{PhysRevD.17.3090} was developed to understand heavy meson spectroscopy.
This model is a non-relativistic approach for heavy quarks with interactions that are governed by $SU(3)$
color gauge symmetry, with flavor only broken by the quark masses. The main pieces of the model are a Coulomb-like interaction,
inspired by the one-gluon exchange, and a linear term, which describes the confining effect. 
It also took
into account two important features than one expects from QCD, Heavy Flavor Symmetry (HFS) and
Heavy Quark Spin Symmetry (HQSS), considering terms that are flavor and spin independent. The naive quark
model from Cornell was fitted to the only 11 states that were known in 1978 and gave a quite good description
of the charmonium and bottomonium spectrum at that time~\cite{PhysRevD.21.203}. Not only that, the predictions
were in quite good agreement with the experiments up to 2003, giving a prediction for 15 new states
in the correct energy range. The Cornell potential has been related with the QCD static potential by 
Sumino~\cite{SUMINO2003173} and more recently with NRQCD up to $N^3LO$~\cite{Mateu:2018zym}.

For the baryon spectra, also very soon, in 1979, quark models developed for the light sector were applied
in the heavy quark sector~\cite{PhysRevD.20.768}. Besides, Stanley and Robsen~\cite{PhysRevLett.45.235} 
extended the Cornell model to study heavy baryons. Many new states were predicted, but it took
a long time to be seen on experiments.

Until 2003, the simple naive quark model picture was in quite good agreement with experiments. However, already
in the original Cornell model~\cite{PhysRevD.17.3090}, the coupling with two-meson thresholds was considered 
for excited states, although was found of no relevance for the states considered at that time. The key event in 2003
for heavy hadron spectroscopy was the discovery of the $X(3872)$ by the Belle 
Collaboration~\cite{PhysRevLett.91.262001}. It was very soon confirmed by the CDF~\cite{PhysRevLett.93.072001}, 
D0~\cite{PhysRevLett.93.162002} and BaBar~\cite{PhysRevD.71.071103} Collaborations. It has some intriguing
properties difficult to understand in the naive quark model picture but easily explained when coupled channel
effects are included. 

Nevertheless, one of the clearest indications that coupled-channel effects have to be considered are the famous pentaquarks
measured by LHCb~\cite{PhysRevLett.115.072001,PhysRevLett.122.222001}. These states are unflavored baryons 
in the region of $4.5$ GeV which rules out a three-light quark baryon interpretation. The only possible 
explanation is a pentaquark with three light quarks and a $c\bar c$ pair. Whether these states are compact
pentaquarks or baryon-meson molecules is a matter of intense debate, although the closeness of these states to
different meson-baryon thresholds is seen as a clear indication of the second possibility.

In this work we will make a brief review of coupled-channel effects in the framework of the quark model,
in the same spirit as the original Cornell model. Thus, we will use naive quark-model states coupled to
two-hadron channels, also built from naive quark-model states. The coupling between these two different sectors
will be obtained using the microscopic $^3P_0$ creation model. The paper is organized as follows: In 
Section~\ref{qmodel} we will give a brief introduction to the naive chiral quark model and how to
calculate the spectrum in this picture. In Section~\ref{3P0model} we will give the basis of the $^3P_0$
model and how to evaluate the transition amplitude. Section~\ref{coupling} will be devoted to present
the formalism to couple one and two-hadron states. In Section~\ref{results} we will show a few examples
in the meson spectrum where such effects are relevant and show some results in the quark model picture.
We will end with some conclusions.

\section{The Naive Chiral Quark Model}
\label{qmodel}

The quark model we use is a constituent quark model based on spontaneous chiral symmetry 
breaking~\cite{MANOHAR1984189}. It was first applied to the light-quark sector~\cite{Fernandez_1993}
and then extended to the heavy sector~\cite{Vijande_2005}.

The main ingredients of the model are the following. The spontaneous chiral symmetry breaking generates two
important effects in the light-quark sector. On one side, the light quarks acquire a dynamical mass that,
at zero momentum, is of the
order of 330 MeV for the $u$ and $d$ quarks, and 550 MeV for the $s$ quark. This dynamical effect has been seen
on the lattice~\cite{PhysRevD.86.014506} and it is the idea introduced phenomenologically in the constituent
quark model. On the other side, it introduces the interaction of light quarks through the exchange of 
pseudo-Goldstone bosons. Another important non-perturbative effect is confinement, seen as the fact that
hadrons are only seen in color singlets. We include it phenomenologically using a linear screened confinement
interaction. This effect has also been observed in lattice QCD where, in quenched QCD, a linear rising of the
energy of two static sources with increasing distance is clearly seen~\cite{BALI20011} and, in unquenched QCD,
this string is broken~\cite{PhysRevD.71.114513} when there is enough energy to produce a quark-antiquark
pair. Finally we introduce QCD perturbative effects through the one-gluon exchange 
interaction~\cite{PhysRevD.12.147}. The model has been review in many works and all the details can be
found in Refs.~\cite{PhysRevD.78.114033,doi:10.1142/S0218301313300269}.

Once the model interaction is settled, in order to obtain the hadron wave functions one has to solve
a non-relativistic bound equation for the two-body problem, in the case of mesons, as quark-antiquark pairs
or the three body problem, for baryons, as three-quark states. There are many different approaches to solve
these systems. The problem can be solved in coordinate space, in momentum
space or using the Raileigh-Ritz variational principle using a certain basis function. Sometimes the potential
involved poses special problems for some technique. For example, the use of a Coulomb-like potential makes the diagonal
part of the potential in momentum space logarithmically divergent, which introduces a numerical problem. 
On the other hand, if we use a non-local interaction in coordinate space, the Schr\"odinger equation ends up
being an integro-differential equation, which is also numerically more demanding. Another important fact is
if coupled channels are considered or not. Momentum space calculations or variational calculations are very
easily extended to such a case, while coordinate space calculations are not straight-forwardly extended.

Nonetheless, the method that can be usually used in any case is the variational method. This method is specially
interesting for our purposes since we will be able to calculate transition amplitudes based on the $^3P_0$ model
with matrix elements of the basis functions used.

The main problem of the method is to find the appropriate basis functions. There are many different options for many
different systems. However, the Gaussian Expansion Method (GEM) has been shown to be a very good approach in almost
any case. The method was firstly proposed by Kamimura~\cite{PhysRevA.38.621} and has been applied to many
few body problems~\cite{HIYAMA2003223,10.1093/ptep/pts015}. 

In the case of the two-body problem, the employed basis functions are a set of Gaussians multiplied by a 
solid-spherical harmonic in the relative coordinate, to take into account the correct behavior of
the wave function at the origin
\begin{eqnarray}
        \varphi_{nlm} (\vec r) &=& N_{nl} r^l e^{-\nu_n r^2} Y_{lm}(\hat r)
        \\
        N_{nl} &=& \left( \frac{2^{l+2} (2\nu_n)^{l+3/2}}{\sqrt \pi (2l+1)!!} \right)^{1/2}
\end{eqnarray}
The basis function is generated by taking several values for the Gaussian parameter $\nu_n$. In the GEM, this
parameters are taken in geometrical progression as
\begin{eqnarray}
        \nu_n &=& \frac{1}{r_n^2}
        \\
        r_n &=& r_{\rm min} a^{n-1}
        \\
        a &=& \bigg( \frac{r_{\rm max}}{r_{\rm min}} \bigg)^{\frac{1}{N-1}}
\end{eqnarray}
where $r_{\rm min}$ and $r_{\rm max}$ are the minimum and maximum radius and $N$ the number of Gaussians.
Numerically, it is important that the parameter $a$ is not too close to 1 so that the problem does not
become singular. Then, one can use the expansion
\begin{eqnarray}
        \varphi (\vec r) &=& \sum_{nl} c_{nl} \varphi_{nlm} (\vec r) 
\end{eqnarray}
and ends up with the generalized eigenvalue problem
\begin{eqnarray}
      \sum_{nl} (  H_{n'l',nl} - E N_{n'l',nl} ) c_{nl} &=& 0
\end{eqnarray}
with
\begin{eqnarray}
      H_{n'l',nl} &=& \langle \varphi_{n'l'm} | H | \varphi_{nlm} \rangle
\\
      N_{n'l',nl} &=& \langle \varphi_{n'l'm} | \varphi_{nlm} \rangle
\end{eqnarray}
In the case of the naive quark model one has to include the spin-flavor-color degrees of freedom so the total
wave function is
\begin{eqnarray}
        \varphi (\vec r) &=& \sum_{nl} c_{nl} \big[ \varphi_{nl} (\vec r) \chi_S \big]_J \chi_F \xi_c
\end{eqnarray}
where $\chi_S$ is the spin wave function, $\chi_F$ is the flavor wave function and $\xi_c$ is the
color singlet quark-antiquark wave function and where
the orbital angular momentum $l$ is coupled with the spin $S$ to total angular momentum $J$.

As a matter of fact, the GEM is more interesting when we have more than two interacting particles. Usually, one chooses
a set of coordinates that includes the center of mass, so one solves for the relative motion of the
interacting particles. Then, one builds the most general wave function with the desired total quantum numbers.
This is usually done considering an expansion in angular momentum of the different coordinates and one
assumes that, for short-range interactions, only the lowest partial waves will be needed. To have a feeling of
what it is needed, typical accurate three-body Fadeev calculations of Triton binding energy needs up to 38 of such
partial waves. The GEM approaches the problem in a different way, it considers also the lowest partial waves
although not only in one set of possible Jacobi coordinates, but in different sets. This has been shown to
have a much faster convergence than the previous approach, which numerically is less demanding because
radial wave functions in a less number of partial waves is needed. The drawback is that now different partial
waves are not orthogonal, so they cannot be considered separately.

In the case of the three-body problem, there are three different sets of Jacobi coordinates given by
\begin{eqnarray}
\vec R_{cm} &=& \frac{m_1 \vec u_1 + m_2 \vec u_2 + m_3 \vec u_3}{M_T}
\\
\vec r_i &=& \vec u_j - \vec u_k
\\
\vec R_i &=& \vec u_i - \frac{m_j \vec u_j + m_k \vec u_k}{m_j+m_k}
\end{eqnarray}
where $\vec u_i$ is the position vector of particle $i$ and $(ijk)$ is one of the 3 even 
permutation of $(123)$. The orbital wave function is taken as
\begin{eqnarray}
        \varphi &=& \sum_{i=1}^3 \sum_{n_il_iN_iL_i} c_{n_il_iN_iL_i} 
\big[ \varphi_{n_il_i} (\vec r_i) \varphi_{N_iL_i} (\vec R_i) \big]_{L_T} 
\end{eqnarray}
where $(-1)^{l_i+L_i}$ gives the parity.
If there are identical particles, some relations between different modes $i$ may be needed. These relations
can be easily obtained from the action of the permutation operator
\begin{eqnarray}
P_{ij} \big[ \varphi_{n_il_i} (\vec r_i) \varphi_{N_iL_i} (\vec R_i) \big]_{L_T} &=&
(-1)^{l_i} \big[ \varphi_{n_il_i} (\vec r_j) \varphi_{N_iL_i} (\vec R_j) \big]_{L_T} 
\\
P_{ij} \big[ \varphi_{n_jl_j} (\vec r_j) \varphi_{N_jL_j} (\vec R_j) \big]_{L_T} &=&
(-1)^{l_j} \big[ \varphi_{n_jl_j} (\vec r_i) \varphi_{N_jL_j} (\vec R_i) \big]_{L_T} 
\\
P_{ij} \big[ \varphi_{n_kl_k} (\vec r_k) \varphi_{N_kL_k} (\vec R_k) \big]_{L_T} &=&
(-1)^{l_k} \big[ \varphi_{n_kl_k} (\vec r_k) \varphi_{N_kL_k} (\vec R_k) \big]_{L_T} 
\end{eqnarray}

If we consider for example the Helium atom, with particle 1 being the nuclei and particles 2 and 3, the electrons
then we can consider two different wave functions with definite symmetry against the $P_{23}$ operator
\begin{eqnarray}
        \varphi &=& \sum_{n_1l_1N_1L_1} c_{n_1l_1N_1L_1} 
\big[ \varphi_{n_1l_1} (\vec r_1) \varphi_{N_1L_1} (\vec R_1) \big]_{L_T} 
\\
        \varphi &=& \sum_{n_{23}l_{23}N_{23}L_{23}} c_{n_{23}l_{23}N_{23}L_{23}} \bigg\{
\big[ \varphi_{n_{23}l_{23}} (\vec r_2) \varphi_{N_{23}L_{23}} (\vec R_2) \big]_{L_T} 
\pm \big[ \varphi_{n_{23}l_{23}} (\vec r_3) \varphi_{N_{23}L_{23}} (\vec R_3) \big]_{L_T} 
\bigg\}
\end{eqnarray}
where the symmetry to the exchange of the electrons is $(-1)^{l_1}$ in the first case and $\pm(-1)^{l_{23}}$
for the second case. Then, the spin wave functions of the electrons have to be considered to solve
the wave functions for parahelium ($S=0$) and orthohelium ($S=1$). In Tables~\ref{parah} and \ref{ortoh}
we show the ground state and first excited states for parahelium and orthohelium, respectively, compared
to experimental data from the NIST database. Here we only include Coulomb interactions so we should
expect deviations of the order of $\alpha^2 \sim 10^{-4}$. As we can see, even with a long-range interaction
as the Coulomb one, the GEM works very well.

\begin{table}
\caption{First levels of parahelium ($S=0$) with orbital angular momentum $L\le 2$. The 
experimental data are taken from NIST, Ref.~\cite{NIST1}. Results with the GEM method from this work.\label{parah}}
\centering
\begin{tabular}{ccccccc}
\toprule
       & \textbf{term} & $J$ & \textbf{NIST} & \textbf{GEM} \\
\midrule
$1s^2$ & $^1S$  & 0 &  0.00000000      &  0.00 \\
\midrule
$1s2s$ & $^1S$  & 0 & 20.6157751334    & 20.61 \\
\midrule  	  	  	  	  	 
$1s2p$ & $^1P$  & 1 & 21.2180230218    & 21.21 \\
\midrule  	  	  	  	  	 
$1s3s$ & $^1S$  & 0 & 22.920317682     & 22.91 \\
\midrule  	  	  	  	  	 
$1s3d$ & $^1D$  & 2 & 23.07407511941   & 23.07 \\
\midrule  	  	  	  	  	 
$1s3p$ & $^1P$  & 1 & 23.0870188528    & 23.08 \\
\midrule  	  	  	  	  	 
$1s4s$ & $^1S$  & 0 & 23.6735709133    & 23.67 \\
\midrule  	  	  	  	  	 
$1s4d$ & $^1D$  & 2 & 23.73633535786   & 23.73 \\
\midrule  	  	  	  	  	 
$1s4p$ & $^1P$  & 1 & 23.7420703918    & 23.74 \\
\midrule  	  	  	  	  	 
$1s5s$ & $^1S$  & 0 & 24.0112153129    & 24.00 \\
\midrule  	  	  	  	  	 
$1s5d$ & $^1D$  & 2 & 24.042803734930  & 24.04 \\
\midrule  	  	  	  	  	 
$1s5p$ & $^1P$  & 1 & 24.0458007297    & 24.04 \\ 
\midrule  	  	  	  	  	 
$1s6s$ & $^1S$  & 0 & 24.1911605982    & 24.18 \\
\midrule  	  	  	  	  	 
$1s6d$ & $^1D$  & 2 & 24.209250116893  & 24.20 \\
\bottomrule
\end{tabular}
\end{table}

\begin{table}
\caption{First levels of orthohelium ($S=1$) with orbital angular momentum $L\le 2$. The 
experimental data are taken from NIST, Ref.~\cite{NIST1}. Results with the GEM method from this work. \label{ortoh}}
\centering
\begin{tabular}{ccccccc}
\toprule
       & \textbf{term} & $J$ & \textbf{NIST} & \textbf{GEM} \\
\midrule
$1s2s$ & $3S$  & 1 & 19.81961484203   & 19.82 \\
\midrule
$1s2p$ & $3P$  & 2 & 20.96408720675   & 20.98 \\
       &       & 1 & 20.96409668230   &       \\
       &       & 0 & 20.96421916817   &       \\
\midrule  	  	  	  	
$1s3s$ & $3S$  & 1 & 22.718466742     & 22.71 \\
\midrule  	  	  	  	  	 
$1s3p$ & $3P$  & 2 & 23.0070734673    & 23.01 \\
       &       & 1 & 23.0070761918    &       \\
       &       & 0 & 23.0071097475    &       \\
\midrule  	  	  	  	  	 
$1s3d$ & $3D$  & 3 & 23.07365102990   & 23.07 \\
       &       & 2 & 23.07365134140   &       \\
       &       & 1 & 23.07365682165   &       \\
\midrule  	  	  	  	  	 
$1s4s$ & $3S$  & 1 & 23.593959036     & 23.59 \\
\midrule  	  	  	  	  	 
$1s4p$ & $3P$  & 2 & 23.7078915511    & 23.71 \\
       &       & 1 & 23.7078926664    &       \\
       &       & 0 & 23.7079063452    &       \\
\midrule  	  	  	  	  	 
$1s4d$ & $3D$  & 3 & 23.73609051247   & 23.73 \\
       &       & 2 & 23.73609066143   &       \\
       &       & 1 & 23.73609295768   &       \\
\midrule  	  	  	  	  	 
$1s5s$ & $3S$  & 1 & 23.9719717413    & 23.97 \\
\midrule  	  	  	  	  	 
$1s5p$ & $3P$  & 2 & 24.0282253870    & 24.02 \\
       &       & 1 & 24.0282259477    &       \\
       &       & 0 & 24.0282328220    &       \\
\midrule  	  	  	  	  	 
$1s5d$ & $3D$  & 3 & 24.042662564819  & 24.04 \\
       &       & 2 & 24.042662644310  &       \\
       &       & 1 & 24.042663817021  &       \\
\midrule  	  	  	  	  	 
$1s6s$ & $3S$  & 1 & 24.1689985463    & 24.16 \\ 
\midrule  	  	  	  	  	 
$1s6p$ & $3P$  & 2 & 24.2008157776    & 24.20 \\ 
       &       & 1 & 24.2008160981    &       \\
       &       & 0 & 24.2008200312    &       \\
\midrule  	  	  	  	  	 
$1s6d$ & $3D$  & 3 & 24.209163433335  & 24.22 \\ 
       &       & 2 & 24.209163480258  &       \\
       &       & 1 & 24.209164158016  &       \\
\bottomrule  	  	  	  	  	 
\end{tabular}
\end{table}
 	
If the three particles are identical then the wave function to be used is
\begin{eqnarray}
        \varphi &=& \sum_{nlNL} c_{nlNL}  \sum_{i=1}^3 
\big[ \varphi_{nl} (\vec r_i) \varphi_{NL} (\vec R_i) \big]_{L_T} 
\end{eqnarray}

As explained above, we have to include the spin-flavor-color wave function where, again, the spin is coupled with the
total orbital angular momentum $L_T$ to give a total angular momentum $J$ and the color wave function
corresponds to the color singlet.

The GEM is again very accurate and in Table~\ref{Bad} we give the result of ground state heavy baryons
in the Bhaduri~\cite{Bhaduri} model, calculated with the GEM and compared with a Fadeev calculation by
B.~Silvestre-Brac~\cite{Silvestre}. The GEM calculation only includes wave functions with $l_i=L_i=0$.
Besides, the matter radius square and the charge
radius square defined by
\begin{eqnarray}
        \langle R_m^2 \rangle &=& \langle \Psi | \sum_{i=1}^3 \frac{m_i}{M} (\vec u_i - \vec R_{cm})^2 | \Psi \rangle
\\
        \langle R_c^2 \rangle &=& \langle \Psi | \sum_{i=1}^3 e_i (\vec u_i - \vec R_{cm})^2 | \Psi \rangle
\end{eqnarray}
are given. From the results, one expects to have a good approximation to the solution of the three-body problem using the GEM.

\begin{table}
\caption{\label{Bad} Results for the Bhaduri potential obtained with a Fadeev calculation (FD)~\cite{Silvestre}
and the GEM method (this work). Masses are given in MeV, while matter radius square $R^2_m$ and charge
radius square $R^2_c$ are in fm$^2$.}
\centering
\begin{tabular}{ccccccc}
\toprule
\textbf{State}               & \textbf{M(FD)}&  \textbf{M(GEM)}  & \textbf{$R_m^2$(FD)}  & \textbf{$R_m^2$(GEM)} & \textbf{$R_c^2$(FD)}  & \textbf{$R_c^2$(GEM)} \\
\midrule
$\Lambda_c^+(cud)$    & 2300 & 2298.5 &   0.097      &  0.0984      &  0.117        &  0.1180      \\
\midrule
$\Sigma_c^0(cud)$     & 2473 & 2475.0 &   0.111      &  0.1116      & -0.224        & -0.2247      \\
$\Sigma_c^+(cud)$     &      &         &              &              &  0.134        &  0.1347      \\
$\Sigma_c^{++}(cud)$  &      &         &              &              &  0.494        &  0.4941      \\
\midrule
$\Lambda_b(bud)$      & 5653 & 5649.6 &   0.043      &  0.0435      &  0.115        &  0.1169      \\
\midrule
$\Sigma_b^-(bud)$     & 5858 & 5859.8 &   0.051      &  0.0509      & -0.280        & -0.2804      \\
$\Sigma_b^0(bud)$     &      &         &              &              &  0.138        &  0.1383      \\
$\Sigma_b^+(bud)$     &      &         &              &              &  0.555        &  0.5571      \\
\midrule
$\Xi^0_{c}(nsc)$      & 2490 & 2490.9 &   0.097      &  0.0978      & -0.145        & -0.1463      \\
$\Xi^+_{c}(nsc)$      &      &         &              &              &  0.160        &  0.1617      \\
\midrule
$\Omega_c^0(css)$     & 2700 & 2701.0 &   0.100      &  0.0999      & -0.111        & -0.1111      \\
\midrule
$\Xi^-_{b}(nsb)$      & 5826 & 5824.8 &   0.045      &  0.0459      & -0.193        & -0.1951      \\
$\Xi^0_{b}(nsb)$      &      &         &              &              &  0.151        &  0.1517      \\
\midrule
$\Omega_b^-(bss)$     & 6046 & 6046.7 &   0.050      &   0.0505     & -0.164        & -0.1642      \\
\midrule
$\Xi^{+}_{cc}(ncc)$   & 3631 & 3632.2 &   0.076      &   0.0766     & -0.034        & -0.0330      \\
$\Xi^{++}_{cc}(ncc)$  &      &         &              &              &  0.285        &  0.2852      \\
\midrule
$\Xi^{-}_{bb}(nbb)$   &10197 &10197.4 &   0.031      &   0.0309     & -0.128        & -0.1279      \\
$\Xi^{0}_{bb}(nbb)$   &      &         &              &              &  0.215        &  0.2140      \\
\midrule
$\Omega^+_{cc}(scc)$  & 3739 & 3738.7 &   0.073      &   0.0739     &  0.008        &  0.0091      \\
\midrule
$\Omega^+_{cb}(scb)$  & 7023 & 7024.2 &   0.043      &   0.0430     & -0.023        & -0.0232      \\
\midrule
$\Omega^-_{bb}(sbb)$  &10271 &10271.3 &   0.030      &   0.0304     & -0.083        & -0.0829      \\
\midrule
$\Omega^{++}_{ccc}(ccc)$ & 4806 & 4807.2 &   0.062      &   0.0619  &  0.124        &  0.1239      \\
\midrule
$\Omega^+_{ccb}(ccb)$ & 8032 & 8030.9 &   0.038      &   0.0378     &  0.089        &  0.0891      \\
\midrule
$\Omega^0_{cbb}(cbb)$ &11220 &11218.6 &   0.026      &   0.0264     &  0.032        &  0.0318      \\
\midrule
$\Omega^-_{bbb}(bbb)$ &14370 &14371.8 &   0.019      &   0.0192     & -0.019        & -0.0192      \\
\bottomrule
\end{tabular}
\end{table}

However, not all the states are below the lowest open threshold. If we again consider the Helium atom,
the first ionization occurs at an energy of $24.587389011$ eV~\cite{PhysRevLett.105.063001}, when
the continuum of a ground state of a ${\rm He}^+$ atom and a free electron starts. As the GEM takes
boundary conditions for bound states, one can still find these bound states as shown in Table~\ref{ortohAT},
although it is more difficult. These states can decay into a ${\rm He}^+ + e^-$ so they are resonances
and can be seen on scattering processes. In the case of hadrons there is a similar situation, however
quarks can not abandon a hadron and a quark-antiquark pair is produced to generate two hadron states.

\begin{table}
\caption{Levels of orthohelium ($S=1$) above the first open threshold. The 
experimental data are taken from NIST, Refs.~\cite{doi:10.1063/1.3253119,PhysRevLett.27.367}. 
Results with the GEM method from this work. \label{ortohAT}}
\centering
\begin{tabular}{ccccccc}
\toprule
       & \textbf{term} & \textbf{NIST} & \textbf{GEM} \\
\midrule
$2s2p$ & $3P$  & 58.311~\cite{doi:10.1063/1.3253119}           & 58.31 \\
\midrule
$2p^2$ & $3P$  & 59.67378~\cite{PhysRevLett.27.367}         & 59.66 \\
\midrule  	  	  	  	
$2p3p$ & $3D$  & 63.120~\cite{doi:10.1063/1.3253119}           &       \\
\midrule  	  	  	  	  	 
$2p3d$ & $3D$  & 63.78658~\cite{doi:10.1063/1.3253119}         &       \\
\midrule  	  	  	  	  	 
$2p3d$ & $3P$  & 64.0719~\cite{doi:10.1063/1.3253119}          &       \\
\bottomrule  	  	  	  	  	 
\end{tabular}
\end{table}

\section{The $^3P_0$ Model}
\label{3P0model}

The quark-pair creation model or $^3P_0$ model is a microscopic model that allows to couple channels with
different number of quarks. The name comes from the fact that a quark-antiquark pair is created with
quantum numbers of the vacuum. It was first proposed by Micu~\cite{Micu:1968mk} and, afterwards, Le~Yaouanc
{\it et al} applied it to the strong decays of mesons~\cite{PhysRevD.8.2223} and
baryons~\cite{PhysRevD.9.1415}. These authors also evaluated strong decay partial widths of
the three charmonium states $\psi(3770)$, $\psi(4040)$ and $\psi(4415)$ within
the same model~\cite{LeYaouanc1977397,LeYaouanc197757}. 

The $^3P_0$ model is usually formulated in terms of the Hamiltonian operator
\begin{equation}
H_{I}=\sqrt{3}\,g_{s}\int d^{3}x \, \bar{\psi}(\vec{x})\psi(\vec{x}),
\label{eq:IH3P0}
\end{equation}
where the only parameter of the model is $g_s$. The factor $\sqrt{3}$ is usually not included but in our case
cancels the color factor in the meson sector, so $g_s$ has the usual definition.

It can be also formulated in terms of a transition operator given by
\begin{equation}
T = -\sqrt{3} \, \sum_{\mu,\nu}\int d^{3}\!p_{\mu}d^{3}\!p_{\nu}
\delta^{(3)}(\vec{p}_{\mu}+\vec{p}_{\nu})\frac{g_{s}}{2m_{\mu}}\sqrt{2^{5}\pi}
\times \left[\mathcal{Y}_{1}\left(\frac{\vec{p}_{\mu}-\vec{p}_{\nu}}{2}
\right)\otimes\left(\frac{1}{2}\frac{1}{2}\right)1\right]_{0}a^{\dagger}_{\mu}
(\vec{p}_{\mu})b^{\dagger}_{\nu}(\vec{p}_{\nu}),
\label{eq:Otransition2}
\end{equation}
where $\mu$ $(\nu)$ are the spin, flavor and color quantum numbers of the
created quark (antiquark). The spin of the quark and antiquark is coupled to
one. The ${\cal Y}_{lm}(\vec{p}\,)=p^{l}Y_{lm}(\hat{p})$ is the solid harmonic
defined as a function of the spherical harmonic.

It is common to give the transition operator in terms of the strength of the 
quark-antiquark pair creation from the vacuum $\gamma$ as in Ref.~\cite{PhysRevD.54.6811}.
The relation is given as
\begin{eqnarray}
\gamma=g_{s}/2m
\end{eqnarray}
being $m$ the mass of the pair created, which is usually a light pair.

\begin{figure}
\centering
\includegraphics[width=10 cm]{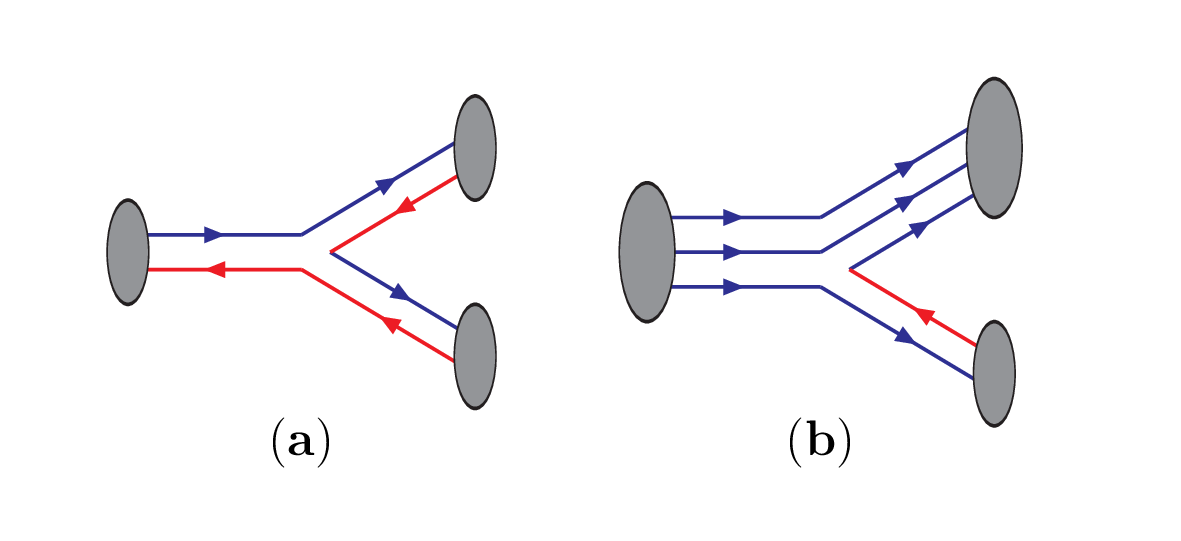}
\caption{Diagrams of the $^3P_0$ model that contributes to the decay of a meson into two mesons (\textbf{a}) and a
baryon into a baryon and a meson (\textbf{b}).}
\end{figure}

We consider a processes where an initial hadron $A$ decays into two final hadrons $B$ and $C$.
When we work in the center of mass system of the initial hadron we have $\vec P_A = \vec P_{cm}=0$,
with $\vec P_A$ the momentum of the initial hadron
and $\vec P_{cm}$ the total momentum of the final hadrons. Then, the matrix element
of the transition operator is written as
\begin{eqnarray}
\langle BC | T | A \rangle = \delta^{(3)}(\vec P_{cm}) {\mathcal M}_{A \to BC}
\end{eqnarray}
The matrix element is taken between hadron states written in terms of quark degrees of freedom in second
quantization. Meson and baryon states are written as
\begin{eqnarray}
|M \rangle &=& \int \phi_M(1,2) b^\dagger_1 d^\dagger_2 |0\rangle
d^3 p_1 d^3 p_2
\\
|B \rangle &=& {\mathcal N}_B \int \Psi_B(1,2,3) b^\dagger_1 b^\dagger_2 b^\dagger_3 |0\rangle
d^3 p_1 d^3 p_2 d^3 p_3 
\end{eqnarray}
where $b^\dagger$ is a quark creation operator and $d^\dagger$ an antiquark creation operator,
${\mathcal N}_B$ is a factor in terms of the number of identical quarks in the baryon, and
with the normalization convention for the wave functions,
\begin{eqnarray}
\sum \int |\phi_M(1,2)|^2 d^3 p_1 d^3 p_2 = 1
\\
\sum \int |\Psi_B(1,2,3)|^2 d^3 p_1 d^3 p_2 d^3 p_3 = 1
\end{eqnarray}
where the sum is for discrete degrees of freedom. With this states, two hadron states with correct quantum
numbers are constructed.

With the transition amplitude, the widths for strong decays can be evaluated
\begin{eqnarray}
\Gamma_{A \to BC} = 2\pi \frac{E_B(k_0)E_C(k_0)}{m_A k_0} |{\mathcal M}_{A\to BC}(k_0)|^2
\end{eqnarray}
with $k_0$ the relative momentum of the two final hadrons. 

The transition amplitude is basically given in terms of the wave function of the naive quark model
considered. There are many factors that are given in terms of the quark model symmetries, but also
an important form factor is given in terms of the overlap of the initial and final hadron wave functions
with the transition operators. The orbital part of the matrix element can be difficult to compute and
this is the reason why the use of the GEM to solve the internal wave function of mesons and baryons
is of special interest. The linearity of the operator allows that, using the expansion of the wave function
in the GEM basis 
\begin{eqnarray}
\left| A \right> = \sum_k c_k \left| A_k \right>
\\
\left| B \right> = \sum_i c_i \left| B_i \right>
\\
\left| C \right> = \sum_j c_j \left| C_j \right>
\end{eqnarray}
one can evaluate the matrix element in terms of matrix elements in the GEM basis as
\begin{eqnarray}
\left< BC | T | A \right> = \sum_{ijk} c^*_i c^*_j c_k 
\left< B_i C_j | T | A_j \right> 
\end{eqnarray}
where the matrix element in the GEM basis $\left< B_i C_j | T | A_j \right>$ are evaluated analytically.

Once the naive quark model is fixed, the only unknown to determine
the transition amplitude is the $^3P_0$ strength parameter $\gamma$. In the case of meson decays,
an scaled dependent parameter was considered in Ref.~\cite{SEGOVIA2012322} as
\begin{eqnarray}
\gamma(\mu) = \frac{\gamma_0}{\log(\frac{\mu}{\mu_0})}
\end{eqnarray}
with $\gamma_0=0.821\pm0.02$ and $\mu_0=(49.84\pm 2.58)$ MeV. The scale is taken as the reduced mass of the
quarks on the initial meson.
The parameters were fixed to the
strong width of a few open-charm, charmonium and bottomonium states and, then, applied to many different
states on these sectors. However it is interesting to notice that this scale dependence was able to
predict the strong decay widths of open-bottom mesons correctly without including this sector on the fit.

\section{The unquenched quark model}
\label{coupling}

In the previous section we showed how one and two-hadron states are connected and give rise to the 
strong decay widths. However, the same transition amplitude has as a consequence: the one-hadron and
two-hadron states connected gets mixed. As the origin is the strong force, this mixing can be
sizable. For this reason, in some cases, it is important to consider the effect and, for this purpose,
we consider the physical state as
\begin{eqnarray}
\left| \Psi \right> = \sum_\alpha c_\alpha \left|\psi_\alpha \right>
 + \sum_\beta \chi_\beta(P)\left|\psi_{H1}\psi_{H2}\beta\right>
\end{eqnarray}
where $|\psi_\alpha\rangle$ are naive quark model one-hadron states with $\alpha$ quantum numbers 
and $|\psi_{H1}\psi_{H2}\beta \rangle$ are two-hadron state $H_1H_2$ with $\beta$ quantum numbers and
relative momentum $P$. We define now the
transition amplitude
\begin{eqnarray}
\langle \psi_{H1}\psi_{H2} \beta| T | \psi_\alpha \rangle = \delta^{(3)}(\vec P_{cm}) P h_{\beta\alpha}(P).
\end{eqnarray}
If we impose the Schr\"odinger equation
\begin{eqnarray}
H \left| \Psi \right> = E \left| \Psi \right> 
\end{eqnarray}
and solve for the one-hadron amplitudes we find
\begin{equation}
c_\alpha = \frac{1}{E-M_\alpha}\sum_\beta \int h_{\alpha\beta}(P)\chi_\beta(P)P^2dP
\end{equation}
where $\chi_\beta(P)$ is given by
the equation in the two-hadron sector
\begin{equation}
\sum_\beta \int \left( H_{\beta' \beta}^{H_1H_2}(P',P) +
V_{\beta'\beta}^{\rm eff}(P',P)\right)\chi_\beta(P) P^2 dP=E\chi_{\beta'}(P')
\end{equation}
Here $H_{\beta' \beta}^{H_1H_2}(P',P)$ is the Hamiltonian generated by the kinetic energy of all the quarks and interaction
between pairs of quarks that depends on the relative momentum of the hadrons, since the other degrees of freedom are
fixed by the hadron states.
There is also a part of the interaction which is generated by the coupling with one-hadron states and
is given by
\begin{equation}
V_{\beta'\beta}^{\rm eff}(P',P) = \sum_\alpha \frac{h_{\beta'\alpha}(P')h_{\alpha\beta}(P)}{E-M_\alpha}
\end{equation}
It is interesting to notice that this effective potential has special relevance at
$E \sim M_\alpha$. Furthermore, one should expect attraction for $E < M_\alpha$ since $V_{\beta\beta}^{eff}(P,P)<0$
and repulsion in the other case. This is important when one considers a certain threshold,
since states above threshold will help to bind a molecule while states below threshold
will help to unbind it. This analysis helps to know when a dynamically generated state can appear.

This formalism is suitable for bound states. However, in order to solve the scattering or consider resonances, it is convenient to 
work with the equivalent Lippmann-Schwinger equation written as
\begin{equation}
T^{\beta'\beta}(E;P',P)=V_T^{\beta'\beta}(P',P)+\sum_{\beta''}\int dP''P''^2 V_T^{\beta'\beta''}(P',P'')\frac{1}{E-E_{\beta''}(P'')}T^{\beta''\beta}(E;P'',P)
\end{equation}
with
\begin{eqnarray}
V_T^{\beta'\beta}(P',P)&=&V^{\beta'\beta}(P',P)+V_{\rm eff}^{\beta'\beta}(P',P)
\\
H_{\beta' \beta}^{H_1H_2}(P',P) &=&
\delta_{\beta'\beta} \frac{\delta(P'-P)}{P^2} \frac{P^2}{2\mu_\beta}
+ V^{\beta'\beta}(P',P)
\end{eqnarray}
The solution to this equations is given in Ref.~\cite{Baru}
\begin{equation}
        T^{\beta'\beta}(E;P',P)={T^{\beta'\beta}_V(E;P',P)}+
        {\sum_{\alpha,\alpha'}\phi^{\beta'\alpha'}(E;P')\Delta^{-1}_{\alpha'\alpha}(E)\bar{\phi}^{\alpha\beta}(E;P)}
\end{equation}
The first term in the right hand side is the non-resonant contribution given by the solution of the equation
\begin{equation}
T_V^{\beta'\beta}(E;P',P)=V^{\beta'\beta}(P',P)+\sum_{\beta''}\int dP''P''^2 
V^{\beta'\beta''}(P',P'')\frac{1}{z-E_{\beta''}(P'')}T_V^{\beta''\beta}(E;P'',P)
\end{equation}
The resonant part include the dressed vertex functions
\begin{eqnarray}
 \phi^{\alpha \beta'}(E;P)&=&h_{\alpha \beta'}(P)-\sum_{\beta}\int  
\dfrac{T^{\beta'\beta}_V(E;P,q)h_{\alpha \beta}(q)}{q^2/2\mu_\beta-E}\,q^2\,dq,\\
 \bar{\phi}^{\alpha \beta}(E;P)&=&h_{\alpha \beta}(P)-\sum_{\beta'}\int 
\dfrac{h_{\alpha \beta'}(q)T^{\beta'\beta}_V(E;q,P)}{q^2/2\mu_{\beta'}-E}\,q^2\,dq
\end{eqnarray}
and the dressed two hadron propagator defined as the inverse of
\begin{eqnarray}
\Delta^{\alpha'\alpha}(E)&=&\left\{(E-M_\alpha)\delta^{\alpha'\alpha}+\mathcal{G}^{\alpha'\alpha}(E)\right\}
\\
\mathcal{G}^{\alpha'\alpha}(E)&=&\sum_\beta \int dq q^2
\dfrac{\phi^{\alpha\beta}(q,E)h_{\beta \alpha'}(q)}{q^2/2\mu_\beta-E}
\end{eqnarray}

The dressed propagator has singularities at the energies of the resonance states so, to find these energies, we solve
the equation
\begin{equation}
         \left|\Delta^{\alpha'\alpha}(\bar E)\right|=\left|(\bar E-M_\alpha)\delta^{\alpha'\alpha}+\mathcal{G}^{\alpha'\alpha}(\bar E)\right|=0.
\end{equation}
Once the resonance energies $\bar E$ are known, we find one-hadron amplitudes by solving
\begin{equation} 
\left\{ M_\alpha\delta^{\alpha\alpha'}-\mathcal{G}^{\alpha'\alpha}(\bar E)\right\}c_{\alpha'}(\bar E)=\bar E\,c_\alpha(\bar E)
\end{equation}
and the two-hadron wave function is given by
\begin{equation} 
\chi_{\beta'}(P')  =  -2\mu_{\beta'}\sum_\alpha\dfrac{\phi_{\beta' \alpha}(E;P')c_\alpha}{P'^2-k^2_{\beta'}}
\end{equation}
Notice that the normalization of the state requires
\begin{equation}
\sum_\alpha|c_\alpha|^2+\sum_\beta\left<\chi_\beta|\chi_\beta\right>=1
\end{equation}
The solution with naive quark model states $|\psi_\alpha\rangle$ is only exact if all the state $\alpha$ are included.
However, including only those states close to the energy range under consideration has been shown to be
a good approximation. In Ref.~\cite{AHEP}, the unquenched quark model for charmonium mesons was considered but solving 
not only for the relative two-hadron wave function, but for the wave function of the $\psi_\alpha$ meson
as a $c\bar c$ meson, getting very similar results to the present approximation.

\section{Coupled channel effects.}
\label{results}

As mentioned before, since 2003 it is clear that the naive quark model is not enough to understand
the heavy hadron spectra. In some cases, as the pentaquarks, the energy scale of its mass make
unavoidable to include higher Fock components. However, as we will see, in other cases
threshold effects can  easily explain
properties very difficult to understand in the naive quark model. In this section we give a few examples
of such cases and we will summarize results obtained using the model previously introduced.

\subsection{Isospin breaking effects.}

The $X(3872)$ was discovered in the $J/\psi\pi\pi$ invariant mass distribution of the
$B^+\to K^+ \pi^+\pi^-J/\psi$ decay. The two-pions in this decay came from a $\rho^0$ meson~\cite{PhysRevLett.96.102002},
which is an isospin 1 final state. The ratio of the decay into three pions was also 
measured
and the three-pions came from the decay of an $\omega$ meson~\cite{Abe:2005ix} which is an isospin 0 channel. The ratio between these two decay modes was found to be
\begin{eqnarray}
        \frac{{\mathcal B}(X(3872)\to \pi^+\pi^-\pi^0 J/\psi)}{{\mathcal B}(X(3872)\to \pi^+\pi^- J/\psi)}=
        1.0 \pm0.4({\rm stat})\pm0.3({\rm syst})
\label{X3872}
\end{eqnarray}
So this state can decay into final states with two different values of isospin, which implies that either isospin
is violated in the decay process or the isospin of the $X(3872)$ is not well defined.

The $X(3872)$ is now included in the PDG as the $\chi_{c1}(3872)$. Quark models usually predict this state at higher
energies, although the deviation can be explained if one considers that this theoretical state is close 
and above the $D\bar D^*$ threshold. One important
question is whether the $X(3872)$ is the state expected in this energy region by quark models, or if it is an
additional state dynamically generated in the $D\bar D^*$ channel. In any case, the crucial property of this state
is that its mass is very close to the $D^0\bar D^{*0}$ with a binding energy given by
\begin{eqnarray}
B \equiv m_{X(3872)}-m_{D^{*0}}-m_{\bar D^0} &=& 1.1^{+0.6+0.1}_{-0.4-0.3}\,
\,{\rm MeV}
\\
&=& (0.00\pm0.18)\,{\rm MeV} 
\\
&=& (0.07\pm0.12)\,{\rm MeV} 
\end{eqnarray}
where the first number is from Ref.~\cite{PhysRevD.81.031103}, the second from
Ref.~\cite{PhysRevLett.122.202002} and the third is from Ref.~\cite{Aaij:2020xjx}.

Since the mass is so close to the $D\bar D^*$ threshold, one would expect a molecule or a mixing with some charmonium
state. Considering the large isospin breaking, the most promising source is the mass splitting between charge and neutral states of $D$ and $D^*$ mesons, finding
\begin{eqnarray}
m_{D^{*+}}+m_{D^-}-m_{D^{*0}}-m_{\bar D^0} = 8.2\pm 0.1\,
\,{\rm MeV} \gg B
\end{eqnarray}
a larger scale than the binding energy, which suggests a big effect. Notice that the masses by themselves
do not suggest it, since the breaking is only of $0.13\,\%$ and $0.08\,\%$ for the $D$ and $D^*$
mesons, respectively. This effect was introduced by Swanson~\cite{SWANSON2004197} in a coupled channel calculation
in which an isospin 1 channel $J/\psi\rho$ was introduced. The important point to notice is that the binding energy for the charged
channel is around $8$ MeV, so the size of this component is around $1.5$ fm, while for the neutral the small
binding energy $B$ makes the size of the order of 4 fm or bigger. This effect generates a big isospin breaking on the
wave function out of the interaction region. This assertion generated some confusion since the isospin breaking
effect is small in the interaction region. The isospin breaking was further analyzed in Ref.~\cite{PhysRevD.80.014003}
in the framework of an Effective Field Theory, were the coupling of the states to the different final channels
could be evaluated. The couplings obtained for $D\bar D^*$ states were $g=2982$ and $g=3005$ for charged and neutral
channels respectively, showing an isospin breaking of less than $1\,\%$. In fact, although the ratio
given in Eq.~\eqref{X3872} suggests a big isospin breaking, this is only due to the big phase space effects 
that enhances the isospin 1 channel against the isospin 0~\cite{PhysRevD.80.014003}. Excluding phase space
effects, the decay in the $\rho$ channel is only around $3.2\,\%$ of the $\omega$ channel decay. This was
clarified in Ref.~\cite{PhysRevD.81.014029} relating the couplings with the probability of the wave function in the
interacting region.

\begin{figure}
\centering
\includegraphics[width=10 cm]{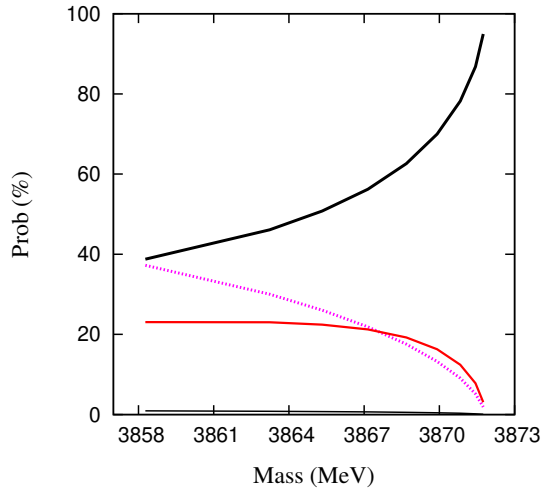}
\caption{\label{fx3872} Probabilities of different components in the physical $X(3872)$ state obtained in the unquenched chiral
quark model as a function of the mass. The mass is varied using the $^3P_0$ strength parameter. 
Lines are $D\bar D^*$, charged channel in black and neutral channel in red, $\chi_{c1}(2P)$ in dashed magenta
and $\chi_{c1}(1P)$ in thin black.}
\end{figure}

The microscopic calculation at the quark level was performed in Ref.~\cite{PhysRevD.81.054023} in the framework
of the chiral quark model previously described. Within the model, the naive quark model $\chi_{c1}(2P)$ state has a mass 
of 3947 MeV, which is far above the $X(3872)$. Coupling with $D\bar D^*$ states makes the naive quark model masses change
slightly. However, the important effect is that a new state appears in the $D\bar D^*$ threshold with properties
in overall good agreement with those of the $X(3872)$. The ratio of Eq.~\eqref{X3872} was analyzed in 
Ref.~\cite{Ortega_2013}, finding a value close to the experimental result. The big isospin breaking on the wave function
is represented in Figure~\ref{fx3872}. For exact isospin symmetry, the $D\bar D^*$ charged and neutral components should have
the same probabilities, while we see that, close to the $D\bar D^*$ neutral threshold, this component dominates.

This effect, seen on the $X(3872)$ state, may appear in any hadron-hadron molecule close to threshold. Of particular
interest are the famous pentaquark states~\cite{PhysRevLett.115.072001,PhysRevLett.122.222001}, which are close to the
$\Sigma_c \bar D$, $\Sigma_c^* \bar D$ and $\Sigma_c \bar D^*$ thresholds. This is the reason why it is widely accepted
that the nature of these states is more likely to be a hadron-hadron molecule than a compact pentaquark state.
Being close to the threshold, isospin breaking effects were studied~\cite{Burns2015} 
and these effects could be magnified in the $P_c(4457)$~\cite{PhysRevD.99.091501}. For this state, the binding energy
is $2.5^{+1.8}_{-4.2}$ MeV for the charged channel with lower threshold and $6.9^{+1.8}_{-4.1}$ MeV for the
higher threshold. Analogously to the decays into $J/\psi\rho$ and $J/\psi\omega$, the pentaquark could decay
into an isospin $3/2$ channel $J/\psi \Delta^+$ or an isospin $1/2$ channel $J/\psi p$. Within an EFT framework, the ratio
\begin{eqnarray}
        R_{\Delta^+/p} = \frac{|{\mathcal B}(P_c(4457)^+\to J/\psi \Delta^+)|}{|{\mathcal B}(P_c(4457)^+\to J/\psi p)|}
        \nonumber
\end{eqnarray}
was evaluated and showed to be up to $30\,\%$. The measurement of this isospin violating decay of the pentaquark could be
the best indication of its molecular nature.

\subsection{HQSS and HFS breaking.}

Heavy Quark Spin Symmetry (HQSS) and Heavy Flavor Symmetry (HFS) are good approximate symmetries of QCD, so 
one would expect them to be realized in the heavy hadron spectrum. 

If we look into the heavy-light sector, under
exact HQSS, the $D$ and $D^*$ mesons should have the same mass. Despite it is not exactly realized,
the ratio $\frac{M_{D^*}-M_D}{M_{D^*}+M_D} \sim 3.6\%$ shows that the breaking is, indeed, small.
In the hidden charm sector, we have $\frac{M_{J/\psi}-M_{\eta_c}}{M_{J/\psi}+M_{\eta_c}} \sim 1.8\%$ and
$\frac{M_{\chi_{c2}(1P)}-M_{\chi_{c0}(1P)}}{M_{\chi_{c2}(1P)}+M_{\chi_{c0}(1P)}} \sim 2.0\%$, so even
smaller breakings. HFS implies that interactions do not depend on the heavy quark mass, so when we find a state
in the charm sector, there must exist an analog in the bottom sector. 

If we consider the $X(3872)$ to be a $D\bar D^*$ molecule, HQSS~\cite{PhysRevD.86.056004,PhysRevD.87.076006,Epel2017}
leads to unavoidable predictions. 
The interaction between $D\bar D^*(1^{++})$ and $D^*\bar D^*(2^{++})$ channels is the same, so if the $X(3872)$ is a $D\bar D^*(1^{++})$ molecule it implies that there
should be a bound state with very similar binding energy in the $D^*\bar D^*(2^{++})$, which was dubbed $X(4012)$.
Besides, HFS requires the interaction between charmed mesons to be the same as for bottom mesons~\cite{PhysRevD.88.054007},
so the same molecules observed in the hidden-charm sector should appear in the hidden-bottom sector.

\begin{figure}
\centering
\includegraphics[width=7 cm]{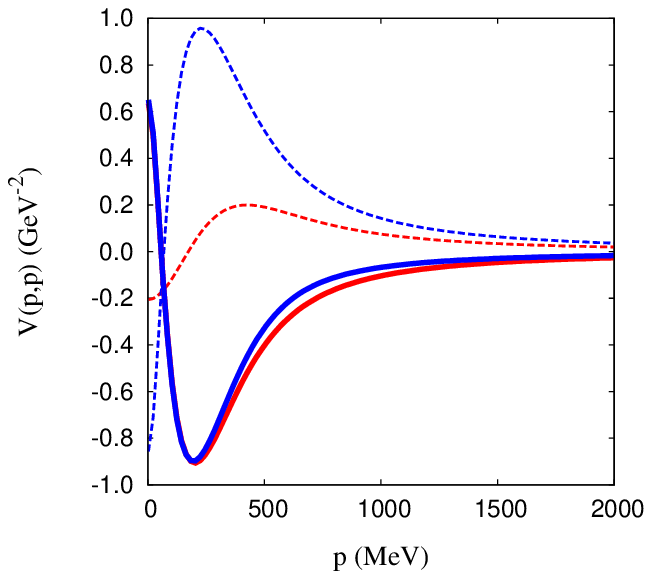}
\includegraphics[width=7 cm]{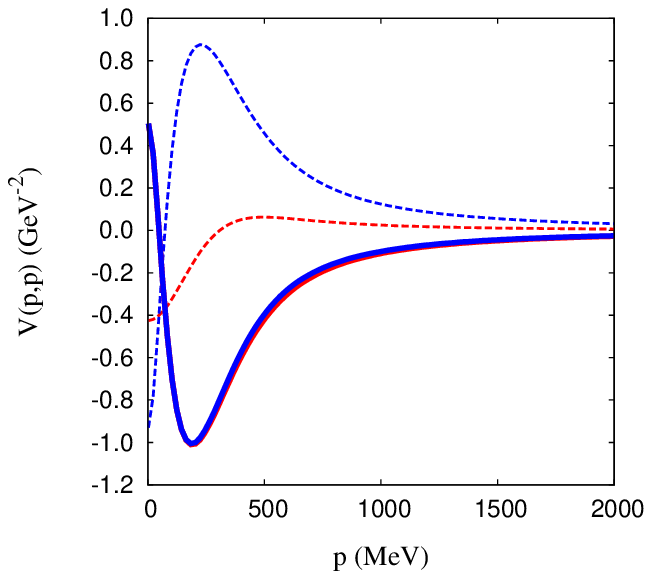}
\caption{\label{fig1} Diagonal matrix elements of the two meson interaction in momentum space for the
  $D^{(*)}\bar D^{(*)}$ sector (left panel)
  and $B^{(*)}\bar B^{(*)}$ sector (right panel).
  For the left panel, the dashed blue line gives the $D^*\bar D^*(0^{++})$ matrix element,
  the dashed red line the $D\bar D(0^{++})$, the solid blue line the right hand side of Eq.(\ref{Ec4}) and the
solid red line left hand side of the same equation. For the right panel, the same but for $B^{(*)}\bar B^ {(*)}$ channels.}
\end{figure}

\begin{figure}
\centering
\includegraphics[width=7 cm]{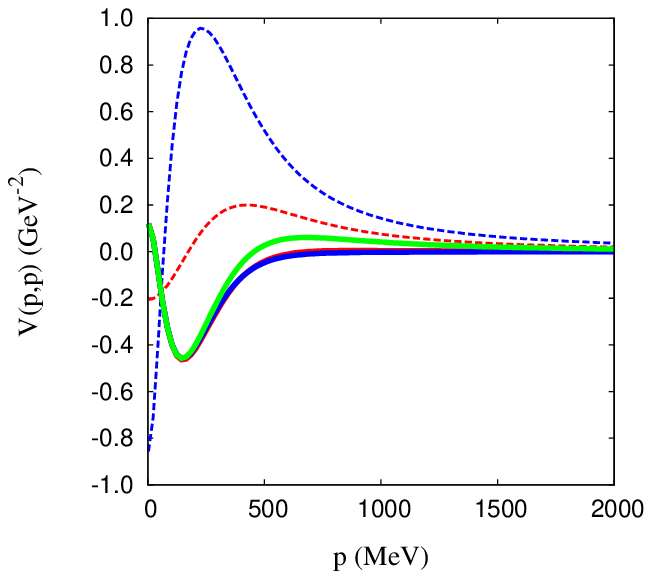}
\includegraphics[width=7 cm]{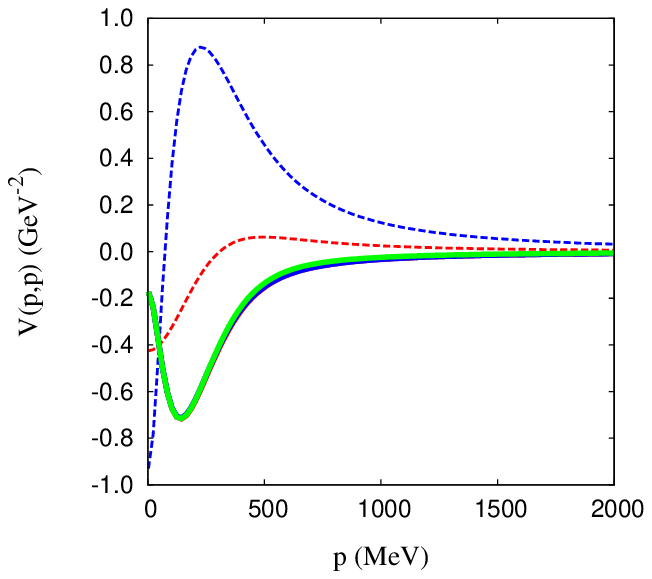}
\caption{\label{fig2} Diagonal matrix elements of the two meson interaction in momentum space for the
  $D^{(*)}\bar D^{(*)}$ sector (left panel)
  and $B^{(*)}\bar B^{(*)}$ sector (right panel).
  For the left panel, the dashed blue line gives the $D^*\bar D^*(0^{++})$ matrix element,
  the dashed red line the $D\bar D(0^{++})$, the solid blue line the right hand side of Eq.(\ref{Ec5}), the
  solid red line the left hand side of the same equation and the solid green line the right hand side of Eq.(\ref{Ec6}). For the right panel, the same but for $B^{(*)}\bar B^ {(*)}$ channels.}
\end{figure}

HQSS is usually fulfilled by heavy quark models, since the heavy quark mass only appears in fine structure
terms that are suppressed as $1/m_Q$ corrections. Within HQSS, one finds for $S$ partial-waves~\cite{doi:10.1063/1.4949442}
\begin{eqnarray}
\frac{2}{\sqrt 3} \langle D^*\bar D^* (0^{++}) | H |D\bar D (0^{++}) \rangle &=&
\langle D\bar D (0^{++}) | H |D\bar D (0^{++}) \rangle
-\langle D^*\bar D^* (0^{++}) | H |D^*\bar D^* (0^{++}) \rangle
\label{Ec4}
\\
\langle D\bar D^* (1^{++}) | H |D\bar D^* (1^{++}) \rangle &=&
\langle D^*\bar D^* (2^{++}) | H |D^*\bar D^* (2^{++}) \rangle
\label{Ec5}
\\ &=&
\frac 3 2
\bigg[
\langle D\bar D (0^{++}) | H |D\bar D (0^{++}) \rangle
 -\frac 1 3 \langle D^*\bar D^* (0^{++}) | H |D^*\bar D^* (0^{++}) \rangle \bigg]
\label{Ec6}
\nonumber \\
\\
2 \langle D\bar D^* (1^{+-}) | H |D\bar D^* (1^{+-}) \rangle &=&
\langle D\bar D (0^{++}) | H |D\bar D (0^{++}) \rangle
+\langle D^*\bar D^* (0^{++}) | H |D^*\bar D^* (0^{++}) \rangle
\end{eqnarray}
where $H$ represents the interacting Hamiltonian.
In the chiral quark model of previous sections, the results for diagonal matrix elements
are shown in Figures~\ref{fig1} and \ref{fig2}, showing that HQSS is approximately fulfilled.
Additionally, comparing the matrix elements of the interactions in the charmed and bottom sectors, we can
see that HFS is also fulfilled.

Nevertheless, open thresholds appear in energy regions where we can also find naive quark model states, so 
one-hadron and two-hadron states can be mixed as shown in the previous sections. This effect can have important
consequences for states that are mainly dynamically generated molecules. Let's see an example and
consider the $P$-wave charmonium states. Considering the allowed spins, using the spectroscopic notation 
$^{2S+1}L_J$, we may have
the states $^1P_1$ ($J^{PC}=1^{+-}$), $^3P_0$ ($0^{++}$), $^3P_1$ ($1^{++}$) and $^3P_2$ 
($2^{++}$) coupled to $^3F_2$. Under HQSS, all these states are degenerated (within a small deviation due to the breakings). 
The largest mass splitting is given by the difference
of the ground states $M_{\chi_{c2}(1P)}-M_{\chi_{c0}(1P)}\sim 141$ MeV, being the splitting for excited
states smaller in the naive quark model picture. The $\chi_{cJ}(2P)$ and $h_c(2P)$ states are in the region of
the $D^{(*)}\bar D^{(*)}$ thresholds, however the threshold difference $M_{D^*\bar D^*}-M_{D\bar D}\sim 282$ is larger. 
Taking only $S$-wave two-meson states, only $D\bar D$ and $D^*\bar D^*$ can have $0^{++}$ quantum numbers,
while only $D\bar D^*$ can have $1^{++}$ and $D^*\bar D^*$ can have $2^{++}$. This means that the relevant threshold
in each channel will have a different relative position with respect to naive quark model states. This is shown
in Fig.~\ref{fcharm}, where we can see that, for the $1^{++}$ channel, the naive quark model state is above the
$D\bar D^*$ threshold, giving additional attraction, while in the $2^{++}$ channel the $P$-wave state is below
the threshold, giving repulsion~\footnote{The state above $D^*\bar D^*$ threshold is an $F$ state.}. This explains why the $1^{++}$ channel has
an additional state, while the $2^{++}$ has not, which is against HQSS expectations.
A systematic study of this effect was performed at hadron level in Ref.~\cite{Cincioglu2016} and
a more elaborate study at the quark level was performed in Ref.~\cite{ORTEGA20181}.

\begin{figure}
\centering
\includegraphics[width=10 cm]{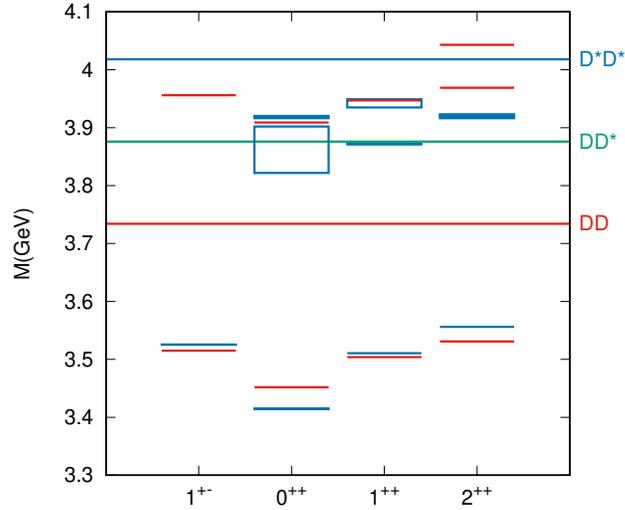}
\caption{\label{fcharm} Charmonium spectrum in the energy region of $1P$ and $2P$ states. Blue boxes
  shows the states in the Particle Data Group~\cite{10.1093/ptep/ptaa104}. 
The $X(3915)$ has been included in the $0^{++}$
and in the $2^{++}$ channels since the $J$ quantum number is not known. The quantum numbers
of the $X(3940)$ are also not known but it has been included
in the $1^{++}$ channel since it has been seen in $D\bar D^*$ but not in $DD$.
  The states
in red are naive $Q\bar Q$ states predicted by the model.}
\end{figure}

In bottomonium we have a different situation. In the $1^{++}$ channel the naive quark model state generates
repulsion and no additional state appears, while in the $2^{++}$ channel there is attraction from the state above
threshold and repulsion from the state below and the final result is that an additional state appears. This
result is, again, against HFS expectations~\cite{doi:10.1063/1.4949442}.
A more elaborate calculation including more thresholds in underway.

\begin{figure}
\centering
\includegraphics[width=10 cm]{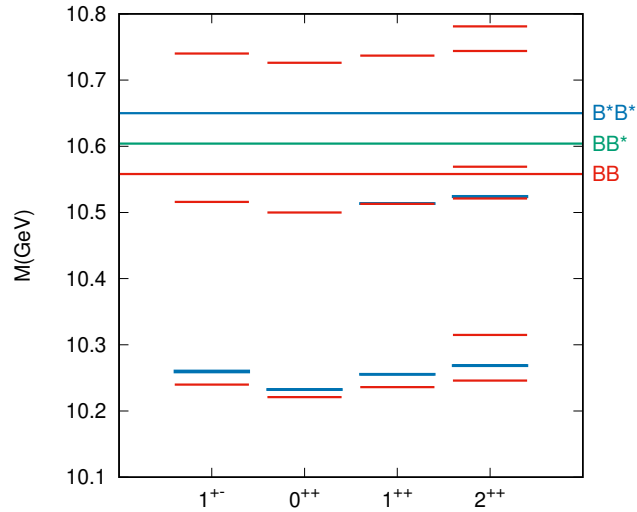}
\caption{\label{fcharm} Bottomonium spectrum in the energy region of $2P$ and $3P$ states. Blue boxes
  shows the states in the Particle Data Group~\cite{10.1093/ptep/ptaa104}. 
  The states in red are the pure $Q\bar Q$ states predicted by the model.}
\end{figure}

\subsection{Threshold cusps}

These are enhancements of the cross sections near the opening of a threshold. One famous example is
the measurement of the scattering lengths difference $a_0-a_2$ in $\pi\pi$ scattering
using a cusp-like enhancement in the $\pi^0\pi^0$ invariant mass distribution in the
$K^\pm \to \pi^\pm \pi^0\pi^0$ decay~\cite{2006173}. It was first noticed in Ref.~\cite{PhysRevLett.6.419}
and then proposed to be measured by Cabibbo~\cite{PhysRevLett.93.121801}. Due to the very precise
experimental data available, a very precised determination of this combination was performed~\cite{Cabibbo_2005}
that was in very good agreement with $\chi$PT predictions.

This effect has been also used to explain strong energy dependencies near threshold of 
invariant mass distributions, as in the case
of $Z_c$ and $Z_b$ states~\cite{PhysRevD.91.034009}. However it has been 
argued~\cite{PhysRevD.91.051504} that such big effects
may not appear without the existence of a nearby pole (bound, virtual or resonance state). 

One example of a threshold cusp effect in the hidden-charm sector is the $X(4140)$ resonance, a
$J^{PC}=1^{++}$ structure observed in the $\phi J/\psi$ invariant mass spectrum by many collaborations, such as CDF~\cite{Aaltonen:2011at}, D0~\cite{Abazov:2015sxa}, 
CMS~\cite{Chatrchyan:2013dma}, Belle~\cite{Shen:2009vs}, 
BaBar~\cite{Lees:2014lra} and LHCb~\cite{Aaij:2016iza}.
Within the chiral quark model described above, a coupled calculation of the main open-charm 
channels~\cite{Ortega:2016hde} showed that the structure just above the $\phi J/\psi$ threshold is not 
caused by the effect of a nearby $c\bar c$ pole, but it is associated to the presence of the $D_s\bar D^*_s$ channel.
The residual $D_s\bar D_s^*$ interaction is strong enough to show a rapid increase of the experimental counts, but
too weak to develop a bound or virtual state.

 Even more interesting is the case of $Z_c$ and $Z_b$ states.
In the chiral quark model the $Z_c$ states have been studied in Ref.~\cite{refId0}. The
$Z_c(3900)^\pm/Z_c(3885)^\pm$ and $Z_c(4020)^\pm$ are meson states in the charmonium
energy range. The fact that they are charged rules out the possibility of being $c\bar c$ states,
and, at least, four quarks are needed. Which is the nature of these charged states is still an open question.

The $Z_c(3900)$ was discovered by the BESIII~\cite{PhysRevLett.110.252001} and 
Belle~\cite{PhysRevLett.110.252002} Collaborations in the $\pi J/\psi$ invariant mass distribution
of the reaction $e^+ e^- \to \pi^+\pi^- J/\psi$. It was then seen by the BESIII 
Collaboration~\cite{PhysRevLett.112.022001} in the $D\bar D^*$ invariant mas distribution of the 
$e^+ e^- \to \pi^\pm (D\bar D^*)^\mp$ reaction with a lower mass and was referred to as the $Z_c(3885)$,
although now are seen as the same state. Soon after this discovery, the BESIII Collaboration reported
the discovery of another charged state, the $Z_c(4020)$, in the reaction 
$e^+e^-\to \pi^+\pi^- h_c$~\cite{PhysRevLett.111.242001}. Later on, also BESIII, reported about
the neutral partner~\cite{PhysRevLett.113.212002}, completing the isospin triplet.

In Ref.~\cite{refId0} $D^{*}\bar D^{(*)}$  was analyzed in the sector $I^G(J^{PC})=1^+(1^{+-})$, within the formalism
previously mentioned. Although here there is no $q\bar q$ state coupled to two-meson components, this system
is interesting for another reason. There are two close-by channels that one would not expect to have an
important effect, the $\pi J/\psi$ and $\rho \eta_c$, since the interactions between these mesons is
expected to be small. However, the non-diagonal interaction $D^{*}\bar D^{(*)}-\pi J/\psi$ and
$D^{*}\bar D^{(*)}-\rho \eta_c$ are dominant, and they do not generate bound states but virtual states.

\begin{figure}[!t]
\includegraphics[width=.5\textwidth]{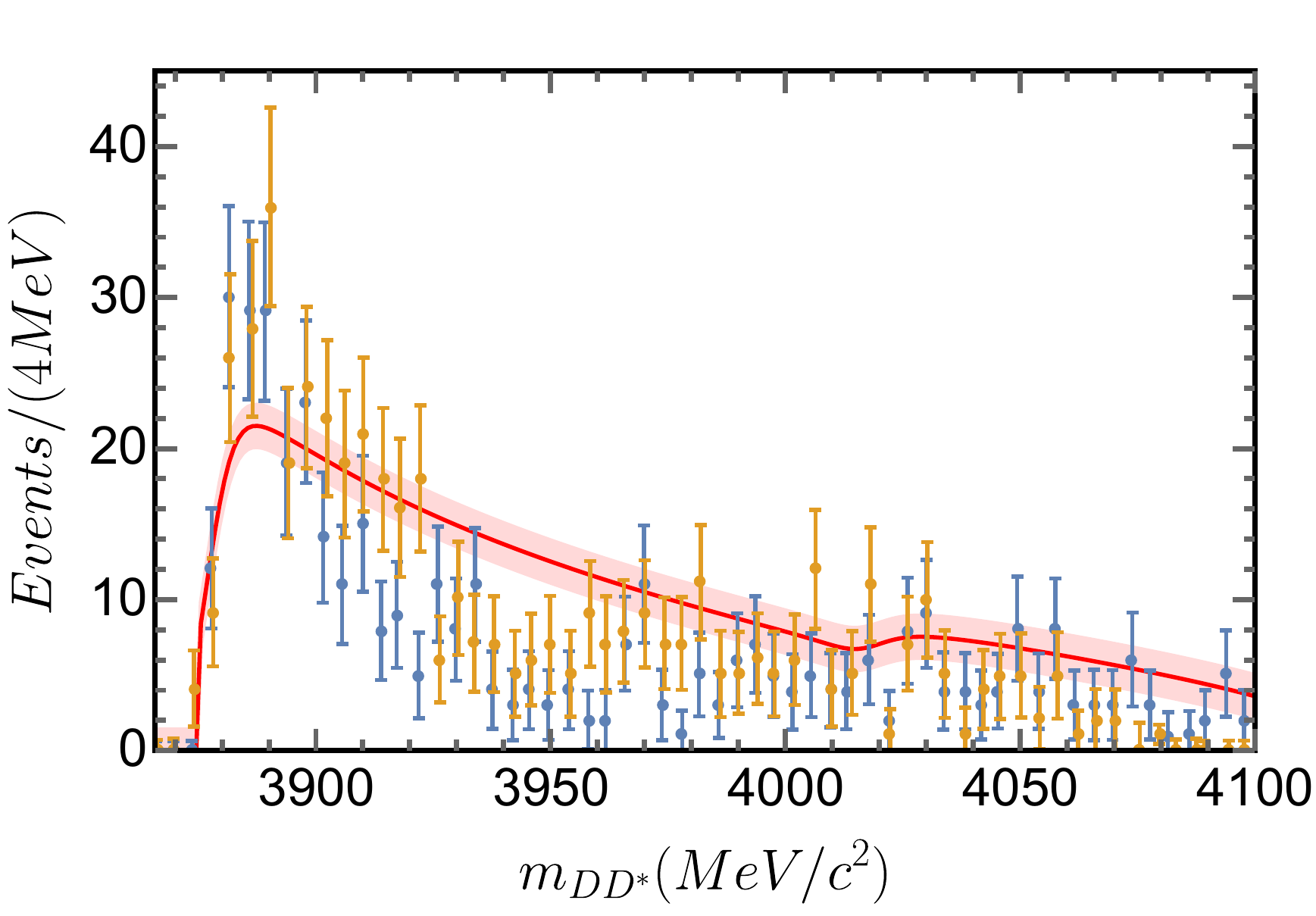}
\includegraphics[width=.5\textwidth]{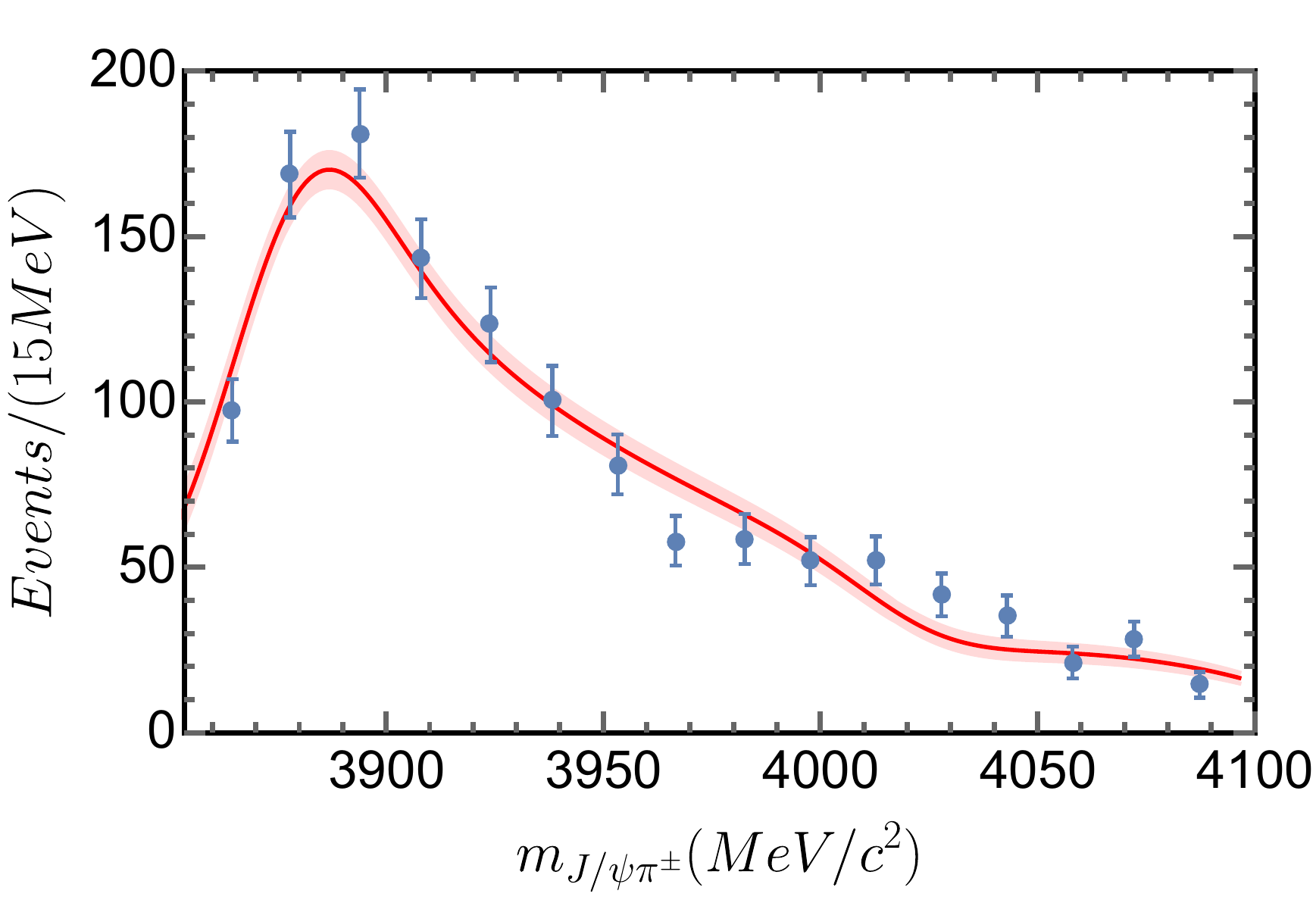} 
\caption{\label{fig:line1} Line shapes for $D\bar D^*$ (left panel) and $\pi J/\psi$ (right panel) at $\sqrt{s}=4.26$ GeV~\cite{refId0}. Experimental data are from Ref.~\cite{Ablikim:2015swa,Collaboration:2017njt}, respectively. The theoretical line shapes have been convoluted with the experimental resolution. The line-shape's $68\%$ uncertainty is shown as a shadowed area.}
\end{figure}

\begin{figure}[!t]
\includegraphics[width=.5\textwidth]{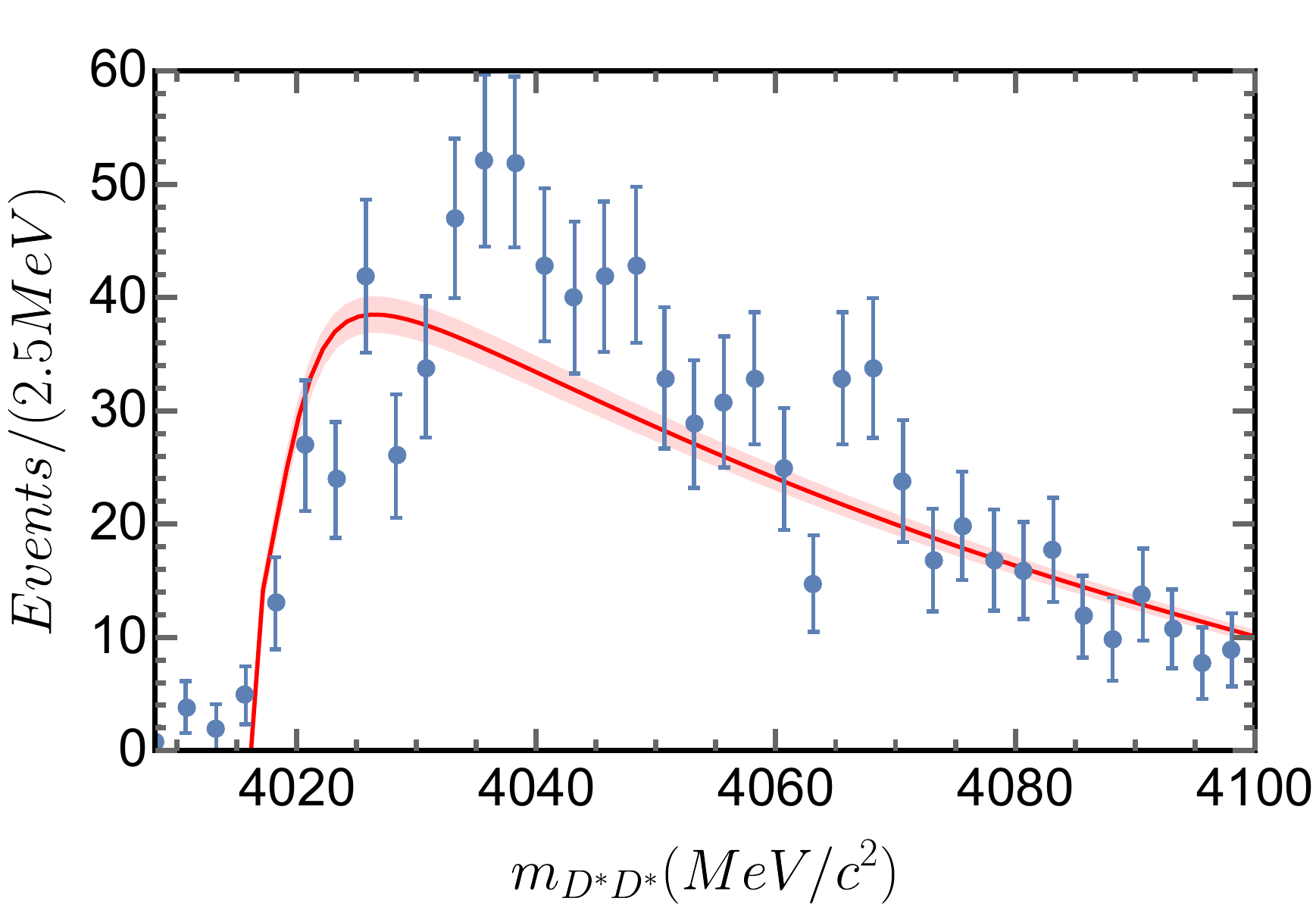}
\includegraphics[width=.5\textwidth]{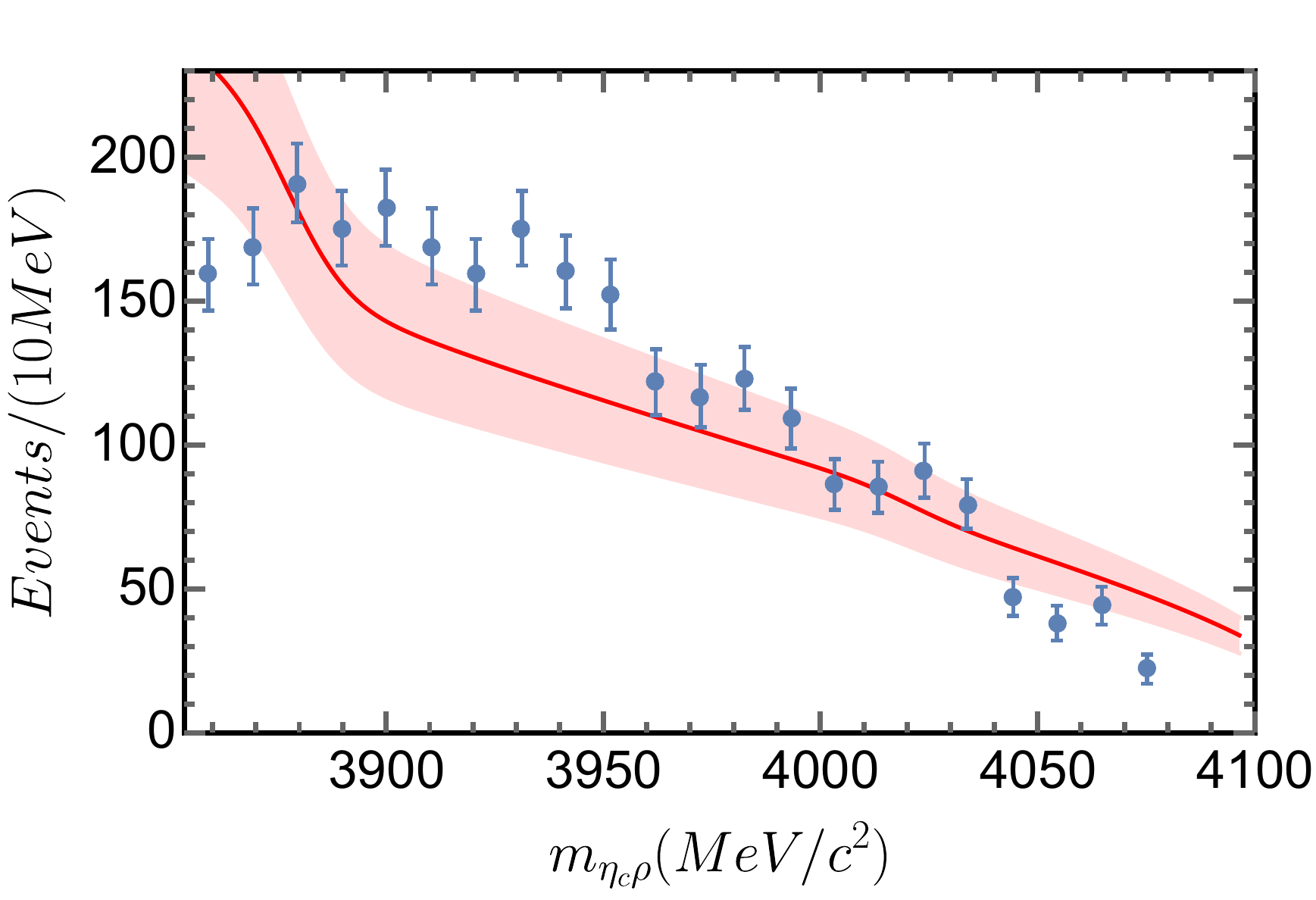}
\caption{\label{fig:line2} Line shapes for $D^*\bar D^*$ (left panel) and $\eta_c\rho$ (right panel) at $\sqrt{s}=4.26$ GeV~\cite{refId0}. Experimental data for $D^*\bar D^*$ are from Refs.~\cite{Ablikim:2013emm}. 
The theoretical line shapes have been convoluted with the experimental resolution. The line-shape's $68\%$-uncertainty is shown as a shadowed area.}
\end{figure}

In Table~\ref{tab:poles} we give the pole position for the states corresponding to the $Z_c(3900)^\pm$ and
$Z_c(4020)^\pm$. The poles are below the $D\bar D^*$ and $D^*\bar D^*$ thresholds, respectively, and in the second Riemann sheet corresponding to virtual
states. Despite these poles already emerge when the main open-charm channels are included, other channels are important
in order to describe the experimental lineshapes. Lineshapes for different reactions are shown in Figures~\ref{fig:line1}
and~\ref{fig:line2}. Although we find poles on the $S$-matrix that produce structures in the lineshapes, in some
cases they don't seem enough, which supports the idea that a threshold cusp effect might not be sufficient to
describe the experimental data without the existence of some associated pole.

\begin{table}
\caption{\label{tab:poles} The $S$-matrix pole positions, in $\text{MeV/c}^2$, for different coupled-channels calculations~\cite{refId0}. The included channels for each case are shown in the first column. Poles are given in the second and fourth columns by the value of the complex energy in a specific Riemann sheet (RS). The RS columns indicate whether the pole has been found in the first (F) or second (S) Riemann sheet of a given channel. Each channel in the coupled-channels calculation is represented as an array's element, ordered with increasing energy.}
\centering
\begin{tabular}{ccccc}
\toprule
Calculation & $Z_c(3900)$ pole & RS & $Z_c(4020)$ pole & RS\\
\midrule
$D\bar D^*$          & $3871.37-2.17\,i$ & (S) & - & - \\  
$D\bar D^*+D^*\bar D^*$  & $3872.27-1.85\,i$ & (S,F) & $4014.16-0.10\,i$ & (S,S) \\  
$\rho\eta_c+D\bar D^*$  & $3871.32-0.00\,i$ & (S,S) & - & - \\   
$\rho\eta_c+D\bar D^*+D^*\bar D^*$  & $3872.07-0.00\,i$ & (S,S,F) & $4013.10-0.00\,i$ & (S,S,S) \\   
$\pi J/\psi+\rho\eta_c+D\bar D^*+D^*\bar D^*$  & $3871.74-0.00\,i$ & (S,S,S,F) & $4013.21-0.00\,i$   & (S,S,S,S) \\
\bottomrule
\end{tabular}
\end{table}

\section{Conclusion}

The naive quark model has been very successful in describing heavy hadron phenomenology for a very long time. However, since 2003,
it seems clear that the coupling with two hadron states are of relevance to describe the phenomenology of new discovered states. 

In this work we have describe how a microscopic quark model can be used, and in particular the chiral quark model, to describe systems
in which conventional quark model states can couple to two-hadron states in a consistent framework. Although these effects are of no
relevance in many low-lying states, keeping the validity of the naive quark model, in some cases the effects can be very important.

This framework have been used during the last years to study the meson spectrum and we have shown a few examples were deviations from 
naive quark model expectations are of special relevance. In fact, threshold effects can generate deviations from expected results  predicted by well-known symmetries such as HFS or HQQS, remarking the importance of analyzing such effects specially in
the heavy meson and baryon sectors.

\vspace{6pt}

\authorcontributions{The authors contributed equally to this work.}

\funding{This work has been funded by Ministerio de
Econom\'\i a, Industria y Competitividad under Contract
No. FPA2016-77177-C2-2-P and Ministerio de Ciencia, Innovaci\'on y Universidades
under Contract No. PID2019-105439GB-C22,
and by EU Horizon 2020 research and innovation programme, STRONG-2020 project, under grant agreement No 824093.}

\conflictsofinterest{The authors declare no conflict of interest.
The funders had no role in the design of the study; in the collection, analyses, or interpretation of data; 
in the writing of the manuscript, or in the decision to publish the results.} 

\abbreviations{The following abbreviations are used in this manuscript:\\

\noindent 
\begin{tabular}{@{}ll}
HQQS & Heavy Quark Spin Symmetry\\
HFS & Heavy Flavor Symmetry\\
$\chi$PT & Chiral Perturbation Theory\\
QCD & Quantum Chromodynamics
\end{tabular}}

\reftitle{References}

\externalbibliography{yes}
\bibliography{Entem}

\begin{thebibliography}{-------}
\providecommand{\natexlab}[1]{#1}

\bibitem[Aubert \em{et~al.}(1974)Aubert, Becker, Biggs, Burger, Chen, Everhart,
  Goldhagen, Leong, McCorriston, Rhoades, Rohde, Ting, Wu, and
  Lee]{PhysRevLett.33.1404}
Aubert, J.J.; Becker, U.; Biggs, P.J.; Burger, J.; Chen, M.; Everhart, G.;
  Goldhagen, P.; Leong, J.; McCorriston, T.; Rhoades, T.G.; Rohde, M.; Ting,
  S.C.C.; Wu, S.L.; Lee, Y.Y.
\newblock Experimental Observation of a Heavy Particle $J$.
\newblock {\em Phys. Rev. Lett.} {\bf 1974}, {\em 33},~1404--1406.
\newblock
  doi:{\changeurlcolor{black}\href{https://doi.org/10.1103/PhysRevLett.33.1404}{\detokenize{10.1103/PhysRevLett.33.1404}}}.

\bibitem[Augustin \em{et~al.}(1974)Augustin, Boyarski, Breidenbach, Bulos,
  Dakin, Feldman, Fischer, Fryberger, Hanson, Jean-Marie, Larsen, L\"uth,
  Lynch, Lyon, Morehouse, Paterson, Perl, Richter, Rapidis, Schwitters,
  Tanenbaum, Vannucci, Abrams, Briggs, Chinowsky, Friedberg, Goldhaber,
  Hollebeek, Kadyk, Lulu, Pierre, Trilling, Whitaker, Wiss, and
  Zipse]{PhysRevLett.33.1406}
Augustin, J.E.; Boyarski, A.M.; Breidenbach, M.; Bulos, F.; Dakin, J.T.;
  Feldman, G.J.; Fischer, G.E.; Fryberger, D.; Hanson, G.; Jean-Marie, B.;
  Larsen, R.R.; L\"uth, V.; Lynch, H.L.; Lyon, D.; Morehouse, C.C.; Paterson,
  J.M.; Perl, M.L.; Richter, B.; Rapidis, P.; Schwitters, R.F.; Tanenbaum,
  W.M.; Vannucci, F.; Abrams, G.S.; Briggs, D.; Chinowsky, W.; Friedberg, C.E.;
  Goldhaber, G.; Hollebeek, R.J.; Kadyk, J.A.; Lulu, B.; Pierre, F.; Trilling,
  G.H.; Whitaker, J.S.; Wiss, J.; Zipse, J.E.
\newblock Discovery of a Narrow Resonance in ${e}^{+}{e}^{\ensuremath{-}}$
  Annihilation.
\newblock {\em Phys. Rev. Lett.} {\bf 1974}, {\em 33},~1406--1408.
\newblock
  doi:{\changeurlcolor{black}\href{https://doi.org/10.1103/PhysRevLett.33.1406}{\detokenize{10.1103/PhysRevLett.33.1406}}}.

\bibitem[Glashow \em{et~al.}(1970)Glashow, Iliopoulos, and
  Maiani]{PhysRevD.2.1285}
Glashow, S.L.; Iliopoulos, J.; Maiani, L.
\newblock Weak Interactions with Lepton-Hadron Symmetry.
\newblock {\em Phys. Rev. D} {\bf 1970}, {\em 2},~1285--1292.
\newblock
  doi:{\changeurlcolor{black}\href{https://doi.org/10.1103/PhysRevD.2.1285}{\detokenize{10.1103/PhysRevD.2.1285}}}.

\bibitem[Herb \em{et~al.}(1977)Herb, Hom, Lederman, Sens, Snyder, Yoh, Appel,
  Brown, Brown, Innes, Ueno, Yamanouchi, Ito, J\"ostlein, Kaplan, and
  Kephart]{PhysRevLett.39.252}
Herb, S.W.; Hom, D.C.; Lederman, L.M.; Sens, J.C.; Snyder, H.D.; Yoh, J.K.;
  Appel, J.A.; Brown, B.C.; Brown, C.N.; Innes, W.R.; Ueno, K.; Yamanouchi, T.;
  Ito, A.S.; J\"ostlein, H.; Kaplan, D.M.; Kephart, R.D.
\newblock Observation of a Dimuon Resonance at 9.5 GeV in 400-GeV
  Proton-Nucleus Collisions.
\newblock {\em Phys. Rev. Lett.} {\bf 1977}, {\em 39},~252--255.
\newblock
  doi:{\changeurlcolor{black}\href{https://doi.org/10.1103/PhysRevLett.39.252}{\detokenize{10.1103/PhysRevLett.39.252}}}.

\bibitem[Kelly \em{et~al.}(1980)Kelly, Horne, Losty, Rittenberg, Shimada,
  Trippe, Wohl, Yost, Barash-Schmidt, Bricman, Dionisi, Mazzucato, Montanet,
  Crawford, Roos, and Armstrong]{RevModPhys.52.S1}
Kelly, R.L.; Horne, C.P.; Losty, M.J.; Rittenberg, A.; Shimada, T.; Trippe,
  T.G.; Wohl, C.G.; Yost, G.P.; Barash-Schmidt, N.; Bricman, C.; Dionisi, C.;
  Mazzucato, M.; Montanet, L.; Crawford, R.L.; Roos, M.; Armstrong, B.
\newblock Review of particle properties.
\newblock {\em Rev. Mod. Phys.} {\bf 1980}, {\em 52},~S1--S286.
\newblock
  doi:{\changeurlcolor{black}\href{https://doi.org/10.1103/RevModPhys.52.S1}{\detokenize{10.1103/RevModPhys.52.S1}}}.

\bibitem[Hagiwara \em{et~al.}(2002)Hagiwara et~al.]{PhysRevD.66.010001}
Hagiwara, K.; others.
\newblock Review of Particle Properties.
\newblock {\em Phys. Rev. D} {\bf 2002}, {\em 66},~010001.
\newblock
  doi:{\changeurlcolor{black}\href{https://doi.org/10.1103/PhysRevD.66.010001}{\detokenize{10.1103/PhysRevD.66.010001}}}.

\bibitem[Zyla \em{et~al.}(2020)Zyla et~al.]{10.1093/ptep/ptaa104}
Zyla, P.A.; others.
\newblock {Review of Particle Physics}.
\newblock {\em Progress of Theoretical and Experimental Physics} {\bf 2020},
  {\em 2020},
  \href{http://xxx.lanl.gov/abs/https://academic.oup.com/ptep/article-pdf/2020/8/083C01/33653179/ptaa104.pdf}{{\normalfont
  [https://academic.oup.com/ptep/article-pdf/2020/8/083C01/33653179/ptaa104.pdf]}}.
\newblock 083C01,
  doi:{\changeurlcolor{black}\href{https://doi.org/10.1093/ptep/ptaa104}{\detokenize{10.1093/ptep/ptaa104}}}.

\bibitem[Cazzoli \em{et~al.}(1975)Cazzoli, Cnops, Connolly, Louttit, Murtagh,
  Palmer, Samios, Tso, and Williams]{PhysRevLett.34.1125}
Cazzoli, E.G.; Cnops, A.M.; Connolly, P.L.; Louttit, R.I.; Murtagh, M.J.;
  Palmer, R.B.; Samios, N.P.; Tso, T.T.; Williams, H.H.
\newblock Evidence for
  $\ensuremath{\Delta}S=\ensuremath{-}\ensuremath{\Delta}Q$ Currents or
  Charmed-Baryon Production by Neutrinos.
\newblock {\em Phys. Rev. Lett.} {\bf 1975}, {\em 34},~1125--1128.
\newblock
  doi:{\changeurlcolor{black}\href{https://doi.org/10.1103/PhysRevLett.34.1125}{\detokenize{10.1103/PhysRevLett.34.1125}}}.

\bibitem[Eichten \em{et~al.}(1978)Eichten, Gottfried, Kinoshita, Lane, and
  Yan]{PhysRevD.17.3090}
Eichten, E.; Gottfried, K.; Kinoshita, T.; Lane, K.D.; Yan, T.M.
\newblock Charmonium: The model.
\newblock {\em Phys. Rev. D} {\bf 1978}, {\em 17},~3090--3117.
\newblock
  doi:{\changeurlcolor{black}\href{https://doi.org/10.1103/PhysRevD.17.3090}{\detokenize{10.1103/PhysRevD.17.3090}}}.

\bibitem[Eichten \em{et~al.}(1980)Eichten, Gottfried, Kinoshita, Lane, and
  Yan]{PhysRevD.21.203}
Eichten, E.; Gottfried, K.; Kinoshita, T.; Lane, K.D.; Yan, T.M.
\newblock Charmonium: Comparison with experiment.
\newblock {\em Phys. Rev. D} {\bf 1980}, {\em 21},~203--233.
\newblock
  doi:{\changeurlcolor{black}\href{https://doi.org/10.1103/PhysRevD.21.203}{\detokenize{10.1103/PhysRevD.21.203}}}.

\bibitem[Sumino(2003)]{SUMINO2003173}
Sumino, Y.
\newblock QCD potential as a “Coulomb-plus-linear” potential.
\newblock {\em Physics Letters B} {\bf 2003}, {\em 571},~173 -- 183.
\newblock
  doi:{\changeurlcolor{black}\href{https://doi.org/https://doi.org/10.1016/j.physletb.2003.05.010}{\detokenize{https://doi.org/10.1016/j.physletb.2003.05.010}}}.

\bibitem[Mateu \em{et~al.}(2019)Mateu, Ortega, Entem, and
  Fern\'andez]{Mateu:2018zym}
Mateu, V.; Ortega, P.G.; Entem, D.R.; Fern\'andez, F.
\newblock {Calibrating the Na\"\i{}ve Cornell Model with NRQCD}.
\newblock {\em Eur. Phys. J. C} {\bf 2019}, {\em 79},~323,
  \href{http://xxx.lanl.gov/abs/1811.01982}{{\normalfont
  [arXiv:hep-ph/1811.01982]}}.
\newblock
  doi:{\changeurlcolor{black}\href{https://doi.org/10.1140/epjc/s10052-019-6808-2}{\detokenize{10.1140/epjc/s10052-019-6808-2}}}.

\bibitem[Copley \em{et~al.}(1979)Copley, Isgur, and Karl]{PhysRevD.20.768}
Copley, L.A.; Isgur, N.; Karl, G.
\newblock Charmed baryons in a quark model with hyperfine interactions.
\newblock {\em Phys. Rev. D} {\bf 1979}, {\em 20},~768--775.
\newblock
  doi:{\changeurlcolor{black}\href{https://doi.org/10.1103/PhysRevD.20.768}{\detokenize{10.1103/PhysRevD.20.768}}}.

\bibitem[Stanley and Robsen(1980)]{PhysRevLett.45.235}
Stanley, D.P.; Robsen, D.
\newblock Do Quarks Interact Pairwise and Satisfy the Color Hypothesis?
\newblock {\em Phys. Rev. Lett.} {\bf 1980}, {\em 45},~235--238.
\newblock
  doi:{\changeurlcolor{black}\href{https://doi.org/10.1103/PhysRevLett.45.235}{\detokenize{10.1103/PhysRevLett.45.235}}}.

\bibitem[Choi \em{et~al.}(2003)Choi et~al.]{PhysRevLett.91.262001}
Choi, S.K.; others.
\newblock Observation of a Narrow Charmoniumlike State in Exclusive
  ${B}^{\ifmmode\pm\else\textpm\fi{}}\ensuremath{\rightarrow}{K}^{\ifmmode\pm\else\textpm\fi{}}{\ensuremath{\pi}}^{+}{\ensuremath{\pi}}^{\ensuremath{-}}J/\ensuremath{\psi}$
  Decays.
\newblock {\em Phys. Rev. Lett.} {\bf 2003}, {\em 91},~262001.
\newblock
  doi:{\changeurlcolor{black}\href{https://doi.org/10.1103/PhysRevLett.91.262001}{\detokenize{10.1103/PhysRevLett.91.262001}}}.

\bibitem[Acosta \em{et~al.}(2004)Acosta et~al.]{PhysRevLett.93.072001}
Acosta, D.; others.
\newblock Observation of the Narrow State
  $X(3872)\ensuremath{\rightarrow}J/\ensuremath{\psi}{\ensuremath{\pi}}^{+}{\ensuremath{\pi}}^{\ensuremath{-}}$
  in $\overline{p}p$ Collisions at $\sqrt{s}=1.96\text{ }\text{
  }\mathrm{T}\mathrm{e}\mathrm{V}$.
\newblock {\em Phys. Rev. Lett.} {\bf 2004}, {\em 93},~072001.
\newblock
  doi:{\changeurlcolor{black}\href{https://doi.org/10.1103/PhysRevLett.93.072001}{\detokenize{10.1103/PhysRevLett.93.072001}}}.

\bibitem[Abazov \em{et~al.}(2004)Abazov et~al.]{PhysRevLett.93.162002}
Abazov, V.M.; others.
\newblock Observation and Properties of the $X(3872)$ Decaying to
  $J/\ensuremath{\psi}{\ensuremath{\pi}}^{+}{\ensuremath{\pi}}^{\ensuremath{-}}$
  in $p\overline{p}$ Collisions at $\sqrt{s}=1.96\text{ }\text{
  }\mathrm{T}\mathrm{e}\mathrm{V}$.
\newblock {\em Phys. Rev. Lett.} {\bf 2004}, {\em 93},~162002.
\newblock
  doi:{\changeurlcolor{black}\href{https://doi.org/10.1103/PhysRevLett.93.162002}{\detokenize{10.1103/PhysRevLett.93.162002}}}.

\bibitem[Aubert \em{et~al.}(2005)Aubert et~al.]{PhysRevD.71.071103}
Aubert, B.; others.
\newblock Study of the
  ${B}^{\ensuremath{-}}\ensuremath{\rightarrow}J/\ensuremath{\psi}{K}^{\ensuremath{-}}{\ensuremath{\pi}}^{+}{\ensuremath{\pi}}^{\ensuremath{-}}$
  decay and measurement of the
  ${B}^{\ensuremath{-}}\ensuremath{\rightarrow}X(3872){K}^{\ensuremath{-}}$
  branching fraction.
\newblock {\em Phys. Rev. D} {\bf 2005}, {\em 71},~071103.
\newblock
  doi:{\changeurlcolor{black}\href{https://doi.org/10.1103/PhysRevD.71.071103}{\detokenize{10.1103/PhysRevD.71.071103}}}.

\bibitem[Aaij \em{et~al.}(2015)Aaij et~al.]{PhysRevLett.115.072001}
Aaij, R.; others.
\newblock Observation of $J/\ensuremath{\psi}p$ Resonances Consistent with
  Pentaquark States in
  ${\mathrm{\ensuremath{\Lambda}}}_{b}^{0}\ensuremath{\rightarrow}J/\ensuremath{\psi}{K}^{\ensuremath{-}}p$
  Decays.
\newblock {\em Phys. Rev. Lett.} {\bf 2015}, {\em 115},~072001.
\newblock
  doi:{\changeurlcolor{black}\href{https://doi.org/10.1103/PhysRevLett.115.072001}{\detokenize{10.1103/PhysRevLett.115.072001}}}.

\bibitem[Aaij \em{et~al.}(2019)Aaij et~al.]{PhysRevLett.122.222001}
Aaij, R.; others.
\newblock Observation of a Narrow Pentaquark State, ${P}_{c}(4312{)}^{+}$, and
  of the Two-Peak Structure of the ${P}_{c}(4450{)}^{+}$.
\newblock {\em Phys. Rev. Lett.} {\bf 2019}, {\em 122},~222001.
\newblock
  doi:{\changeurlcolor{black}\href{https://doi.org/10.1103/PhysRevLett.122.222001}{\detokenize{10.1103/PhysRevLett.122.222001}}}.

\bibitem[Manohar and Georgi(1984)]{MANOHAR1984189}
Manohar, A.; Georgi, H.
\newblock Chiral quarks and the non-relativistic quark model.
\newblock {\em Nuclear Physics B} {\bf 1984}, {\em 234},~189 -- 212.
\newblock
  doi:{\changeurlcolor{black}\href{https://doi.org/https://doi.org/10.1016/0550-3213(84)90231-1}{\detokenize{https://doi.org/10.1016/0550-3213(84)90231-1}}}.

\bibitem[Fernandez \em{et~al.}(1993)Fernandez, Valcarce, Straub, and
  Faessler]{Fernandez_1993}
Fernandez, F.; Valcarce, A.; Straub, U.; Faessler, A.
\newblock The nucleon-nucleon interaction in terms of quark degrees of freedom.
\newblock {\em Journal of Physics G: Nuclear and Particle Physics} {\bf 1993},
  {\em 19},~2013--2026.
\newblock
  doi:{\changeurlcolor{black}\href{https://doi.org/10.1088/0954-3899/19/12/007}{\detokenize{10.1088/0954-3899/19/12/007}}}.

\bibitem[Vijande \em{et~al.}(2005)Vijande, Fern{\'{a}}ndez, and
  Valcarce]{Vijande_2005}
Vijande, J.; Fern{\'{a}}ndez, F.; Valcarce, A.
\newblock Constituent quark model study of the meson spectra.
\newblock {\em Journal of Physics G: Nuclear and Particle Physics} {\bf 2005},
  {\em 31},~481--506.
\newblock
  doi:{\changeurlcolor{black}\href{https://doi.org/10.1088/0954-3899/31/5/017}{\detokenize{10.1088/0954-3899/31/5/017}}}.

\bibitem[Burgio \em{et~al.}(2012)Burgio, Schr\"ock, Reinhardt, and
  Quandt]{PhysRevD.86.014506}
Burgio, G.; Schr\"ock, M.; Reinhardt, H.; Quandt, M.
\newblock Running mass, effective energy, and confinement: The lattice quark
  propagator in Coulomb gauge.
\newblock {\em Phys. Rev. D} {\bf 2012}, {\em 86},~014506.
\newblock
  doi:{\changeurlcolor{black}\href{https://doi.org/10.1103/PhysRevD.86.014506}{\detokenize{10.1103/PhysRevD.86.014506}}}.

\bibitem[Bali(2001)]{BALI20011}
Bali, G.S.
\newblock QCD forces and heavy quark bound states.
\newblock {\em Physics Reports} {\bf 2001}, {\em 343},~1 -- 136.
\newblock
  doi:{\changeurlcolor{black}\href{https://doi.org/https://doi.org/10.1016/S0370-1573(00)00079-X}{\detokenize{https://doi.org/10.1016/S0370-1573(00)00079-X}}}.

\bibitem[Bali \em{et~al.}(2005)Bali, Neff, D\"ussel, Lippert, and
  Schilling]{PhysRevD.71.114513}
Bali, G.S.; Neff, H.; D\"ussel, T.; Lippert, T.; Schilling, K.
\newblock Observation of string breaking in QCD.
\newblock {\em Phys. Rev. D} {\bf 2005}, {\em 71},~114513.
\newblock
  doi:{\changeurlcolor{black}\href{https://doi.org/10.1103/PhysRevD.71.114513}{\detokenize{10.1103/PhysRevD.71.114513}}}.

\bibitem[De~R\'ujula \em{et~al.}(1975)De~R\'ujula, Georgi, and
  Glashow]{PhysRevD.12.147}
De~R\'ujula, A.; Georgi, H.; Glashow, S.L.
\newblock Hadron masses in a gauge theory.
\newblock {\em Phys. Rev. D} {\bf 1975}, {\em 12},~147--162.
\newblock
  doi:{\changeurlcolor{black}\href{https://doi.org/10.1103/PhysRevD.12.147}{\detokenize{10.1103/PhysRevD.12.147}}}.

\bibitem[Segovia \em{et~al.}(2008)Segovia, Yasser, Entem, and
  Fern\'andez]{PhysRevD.78.114033}
Segovia, J.; Yasser, A.M.; Entem, D.R.; Fern\'andez, F.
\newblock ${J}^{PC}={1}^{--}$ hidden charm resonances.
\newblock {\em Phys. Rev. D} {\bf 2008}, {\em 78},~114033.
\newblock
  doi:{\changeurlcolor{black}\href{https://doi.org/10.1103/PhysRevD.78.114033}{\detokenize{10.1103/PhysRevD.78.114033}}}.

\bibitem[SEGOVIA \em{et~al.}(2013)SEGOVIA, ENTEM, FERNANDEZ, and
  HERNANDEZ]{doi:10.1142/S0218301313300269}
SEGOVIA, J.; ENTEM, D.R.; FERNANDEZ, F.; HERNANDEZ, E.
\newblock CONSTITUENT QUARK MODEL DESCRIPTION OF CHARMONIUM PHENOMENOLOGY.
\newblock {\em International Journal of Modern Physics E} {\bf 2013}, {\em
  22},~1330026,
  \href{http://xxx.lanl.gov/abs/https://doi.org/10.1142/S0218301313300269}{{\normalfont
  [https://doi.org/10.1142/S0218301313300269]}}.
\newblock
  doi:{\changeurlcolor{black}\href{https://doi.org/10.1142/S0218301313300269}{\detokenize{10.1142/S0218301313300269}}}.

\bibitem[Kamimura(1988)]{PhysRevA.38.621}
Kamimura, M.
\newblock Nonadiabatic coupled-rearrangement-channel approach to muonic
  molecules.
\newblock {\em Phys. Rev. A} {\bf 1988}, {\em 38},~621--624.
\newblock
  doi:{\changeurlcolor{black}\href{https://doi.org/10.1103/PhysRevA.38.621}{\detokenize{10.1103/PhysRevA.38.621}}}.

\bibitem[Hiyama \em{et~al.}(2003)Hiyama, Kino, and Kamimura]{HIYAMA2003223}
Hiyama, E.; Kino, Y.; Kamimura, M.
\newblock Gaussian expansion method for few-body systems.
\newblock {\em Progress in Particle and Nuclear Physics} {\bf 2003}, {\em
  51},~223 -- 307.
\newblock
  doi:{\changeurlcolor{black}\href{https://doi.org/https://doi.org/10.1016/S0146-6410(03)90015-9}{\detokenize{https://doi.org/10.1016/S0146-6410(03)90015-9}}}.

\bibitem[Hiyama(2012)]{10.1093/ptep/pts015}
Hiyama, E.
\newblock {Gaussian expansion method for few-body systems and its applications
  to atomic and nuclear physics}.
\newblock {\em Progress of Theoretical and Experimental Physics} {\bf 2012},
  {\em 2012},
  \href{http://xxx.lanl.gov/abs/https://academic.oup.com/ptep/article-pdf/2012/1/01A204/4459080/pts015.pdf}{{\normalfont
  [https://academic.oup.com/ptep/article-pdf/2012/1/01A204/4459080/pts015.pdf]}}.
\newblock 01A204,
  doi:{\changeurlcolor{black}\href{https://doi.org/10.1093/ptep/pts015}{\detokenize{10.1093/ptep/pts015}}}.

\bibitem[Morton \em{et~al.}(2006)Morton, Wu, and G.W.F.]{NIST1}
Morton, D.; Wu, Q.; G.W.F., D.
\newblock Energy Levels for the Stable Isotopes of Atomic Helium (4He I and 3He
  I).
\newblock {\em Can. J. Phys.} {\bf 2006}, {\em 84},~83.
\newblock
  doi:{\changeurlcolor{black}\href{https://doi.org/10.1139/P06-009}{\detokenize{10.1139/P06-009}}}.

\bibitem[Bhaduri \em{et~al.}(1981)Bhaduri, Cohler, and Nogami]{Bhaduri}
Bhaduri, R.K.; Cohler, L.E.; Nogami, Y.
\newblock {A unified potential for mesons and baryons}.
\newblock {\em Il Nuovo Cimento A} {\bf 1981}, {\em 65}.
\newblock
  doi:{\changeurlcolor{black}\href{https://doi.org/10.1007/BF02827441}{\detokenize{10.1007/BF02827441}}}.

\bibitem[Silvestre-Brac(1996)]{Silvestre}
Silvestre-Brac, B.
\newblock {Spectrum and Static Properties of Heavy Baryons}.
\newblock {\em Few-Body Systems} {\bf 1996}, {\em 20}.
\newblock
  doi:{\changeurlcolor{black}\href{https://doi.org/10.1007/s006010050028}{\detokenize{10.1007/s006010050028}}}.

\bibitem[Kandula \em{et~al.}(2010)Kandula, Gohle, Pinkert, Ubachs, and
  Eikema]{PhysRevLett.105.063001}
Kandula, D.Z.; Gohle, C.; Pinkert, T.J.; Ubachs, W.; Eikema, K.S.E.
\newblock Extreme Ultraviolet Frequency Comb Metrology.
\newblock {\em Phys. Rev. Lett.} {\bf 2010}, {\em 105},~063001.
\newblock
  doi:{\changeurlcolor{black}\href{https://doi.org/10.1103/PhysRevLett.105.063001}{\detokenize{10.1103/PhysRevLett.105.063001}}}.

\bibitem[Martin(1973)]{doi:10.1063/1.3253119}
Martin, W.C.
\newblock Energy Levels of Neutral Helium (4He I).
\newblock {\em Journal of Physical and Chemical Reference Data} {\bf 1973},
  {\em 2},~257--266,
  \href{http://xxx.lanl.gov/abs/https://doi.org/10.1063/1.3253119}{{\normalfont
  [https://doi.org/10.1063/1.3253119]}}.
\newblock
  doi:{\changeurlcolor{black}\href{https://doi.org/10.1063/1.3253119}{\detokenize{10.1063/1.3253119}}}.

\bibitem[Tech and Ward(1971)]{PhysRevLett.27.367}
Tech, J.L.; Ward, J.F.
\newblock Accurate Wavelength Measurement of the $1s2p
  \,^{3}P^{0}-2{p}^{2}\,^{3}P$ Transition in $^{4}\mathrm{He}$ I.
\newblock {\em Phys. Rev. Lett.} {\bf 1971}, {\em 27},~367--370.
\newblock
  doi:{\changeurlcolor{black}\href{https://doi.org/10.1103/PhysRevLett.27.367}{\detokenize{10.1103/PhysRevLett.27.367}}}.

\bibitem[Micu(1969)]{Micu:1968mk}
Micu, L.
\newblock {Decay rates of meson resonances in a quark model}.
\newblock {\em Nucl. Phys.} {\bf 1969}, {\em B10},~521--526.
\newblock
  doi:{\changeurlcolor{black}\href{https://doi.org/10.1016/0550-3213(69)90039-X}{\detokenize{10.1016/0550-3213(69)90039-X}}}.

\bibitem[Le~Yaouanc \em{et~al.}(1973)Le~Yaouanc, Oliver, P\`ene, and
  Raynal]{PhysRevD.8.2223}
Le~Yaouanc, A.; Oliver, L.; P\`ene, O.; Raynal, J.C.
\newblock "Naive" Quark-Pair-Creation Model of Strong-Interaction Vertices.
\newblock {\em Phys. Rev. D} {\bf 1973}, {\em 8},~2223--2234.
\newblock
  doi:{\changeurlcolor{black}\href{https://doi.org/10.1103/PhysRevD.8.2223}{\detokenize{10.1103/PhysRevD.8.2223}}}.

\bibitem[Le~Yaouanc \em{et~al.}(1974)Le~Yaouanc, Oliver, P\`ene, and
  Raynal]{PhysRevD.9.1415}
Le~Yaouanc, A.; Oliver, L.; P\`ene, O.; Raynal, J.C.
\newblock Naive quark-pair---creation model and baryon decays.
\newblock {\em Phys. Rev. D} {\bf 1974}, {\em 9},~1415--1419.
\newblock
  doi:{\changeurlcolor{black}\href{https://doi.org/10.1103/PhysRevD.9.1415}{\detokenize{10.1103/PhysRevD.9.1415}}}.

\bibitem[Yaouanc \em{et~al.}(1977{\natexlab{a}})Yaouanc, Oliver, Pene, and
  Raynal]{LeYaouanc1977397}
Yaouanc, A.L.; Oliver, L.; Pene, O.; Raynal, J.C.
\newblock Strong decays of $\psi$(4028) as a radial excitation of charmonium.
\newblock {\em Physics Letters B} {\bf 1977}, {\em 71},~397--399.
\newblock
  doi:{\changeurlcolor{black}\href{https://doi.org/10.1016/0370-2693(77)90250-7}{\detokenize{10.1016/0370-2693(77)90250-7}}}.

\bibitem[Yaouanc \em{et~al.}(1977{\natexlab{b}})Yaouanc, Oliver, P{\`e}ne, and
  Raynal]{LeYaouanc197757}
Yaouanc, A.L.; Oliver, L.; P{\`e}ne, O.; Raynal, J.
\newblock Why is $\psi$(4414) so narrow?
\newblock {\em Physics Letters B} {\bf 1977}, {\em 72},~57--61.
\newblock
  doi:{\changeurlcolor{black}\href{https://doi.org/10.1016/0370-2693(77)90062-4}{\detokenize{10.1016/0370-2693(77)90062-4}}}.

\bibitem[Ackleh \em{et~al.}(1996)Ackleh, Barnes, and Swanson]{PhysRevD.54.6811}
Ackleh, E.S.; Barnes, T.; Swanson, E.S.
\newblock On the mechanism of open-flavor strong decays.
\newblock {\em Phys. Rev. D} {\bf 1996}, {\em 54},~6811--6829.
\newblock
  doi:{\changeurlcolor{black}\href{https://doi.org/10.1103/PhysRevD.54.6811}{\detokenize{10.1103/PhysRevD.54.6811}}}.

\bibitem[Segovia \em{et~al.}(2012)Segovia, Entem, and
  Fernández]{SEGOVIA2012322}
Segovia, J.; Entem, D.; Fernández, F.
\newblock Scaling of the P03 strength in heavy meson strong decays.
\newblock {\em Physics Letters B} {\bf 2012}, {\em 715},~322 -- 327.
\newblock
  doi:{\changeurlcolor{black}\href{https://doi.org/https://doi.org/10.1016/j.physletb.2012.08.005}{\detokenize{https://doi.org/10.1016/j.physletb.2012.08.005}}}.

\bibitem[Baru \em{et~al.}(2010)Baru, Hanhart, Kalashnikova, Kudryavtsev, and
  Nefediev]{Baru}
Baru, V.; Hanhart, C.; Kalashnikova, Y.S.; Kudryavtsev, A.E.; Nefediev, A.V.
\newblock Interplay of quark and meson degrees of freedom in a near-threshold
  resonance.
\newblock {\em The European Physical Journal A} {\bf 2010}, {\em 44},~93.
\newblock
  doi:{\changeurlcolor{black}\href{https://doi.org/10.1140/epja/i2010-10929-7}{\detokenize{10.1140/epja/i2010-10929-7}}}.

\bibitem[Ortega \em{et~al.}(2019)Ortega, Entem, and Fernández]{AHEP}
Ortega, P.G.; Entem, D.R.; Fernández, F.
\newblock Unquenching the Quark Model in a Nonperturbative Scheme.
\newblock {\em Advances in High Energy Physics} {\bf 2019}, {\em
  2019},~3465159.
\newblock
  doi:{\changeurlcolor{black}\href{https://doi.org/10.1155/2019/3465159}{\detokenize{10.1155/2019/3465159}}}.

\bibitem[Abulencia \em{et~al.}(2006)Abulencia et~al.]{PhysRevLett.96.102002}
Abulencia, A.; others.
\newblock Measurement of the Dipion Mass Spectrum in
  $X(3872)\ensuremath{\rightarrow}J/\ensuremath{\psi}{\ensuremath{\pi}}^{+}{\ensuremath{\pi}}^{\ensuremath{-}}$
  Decays.
\newblock {\em Phys. Rev. Lett.} {\bf 2006}, {\em 96},~102002.
\newblock
  doi:{\changeurlcolor{black}\href{https://doi.org/10.1103/PhysRevLett.96.102002}{\detokenize{10.1103/PhysRevLett.96.102002}}}.

\bibitem[Abe \em{et~al.}(2005)Abe et~al.]{Abe:2005ix}
Abe, K.; others.
\newblock {Evidence for X(3872) ---> gamma J / psi and the sub-threshold decay
  X(3872) ---> omega J / psi}.
\newblock  {Lepton and photon interactions at high energies. Proceedings, 22nd
  International Symposium, LP 2005, Uppsala, Sweden, June 30-July 5, 2005},
  2005,  \href{http://xxx.lanl.gov/abs/hep-ex/0505037}{{\normalfont
  [arXiv:hep-ex/hep-ex/0505037]}}.

\bibitem[Aushev \em{et~al.}(2010)Aushev et~al.]{PhysRevD.81.031103}
Aushev, T.; others.
\newblock Study of the
  $B\ensuremath{\rightarrow}X(3872)(\ensuremath{\rightarrow}{D}^{*0}{\overline{D}}^{0})K$
  decay.
\newblock {\em Phys. Rev. D} {\bf 2010}, {\em 81},~031103.
\newblock
  doi:{\changeurlcolor{black}\href{https://doi.org/10.1103/PhysRevD.81.031103}{\detokenize{10.1103/PhysRevD.81.031103}}}.

\bibitem[Guo(2019)]{PhysRevLett.122.202002}
Guo, F.K.
\newblock Novel Method for Precisely Measuring the $X(3872)$ Mass.
\newblock {\em Phys. Rev. Lett.} {\bf 2019}, {\em 122},~202002.
\newblock
  doi:{\changeurlcolor{black}\href{https://doi.org/10.1103/PhysRevLett.122.202002}{\detokenize{10.1103/PhysRevLett.122.202002}}}.

\bibitem[Aaij \em{et~al.}(2020)Aaij et~al.]{Aaij:2020xjx}
Aaij, R.; others.
\newblock {Study of the $\psi_2(3823)$ and $\chi_{c1}(3872)$ states in $B^+
  \rightarrow \left( J\psi\pi^+\pi^-\right)K^+$ decays}.
\newblock {\em JHEP} {\bf 2020}, {\em 08},~123,
  \href{http://xxx.lanl.gov/abs/2005.13422}{{\normalfont
  [arXiv:hep-ex/2005.13422]}}.
\newblock
  doi:{\changeurlcolor{black}\href{https://doi.org/10.1007/JHEP08(2020)123}{\detokenize{10.1007/JHEP08(2020)123}}}.

\bibitem[Swanson(2004)]{SWANSON2004197}
Swanson, E.S.
\newblock Diagnostic decays of the X(3872).
\newblock {\em Physics Letters B} {\bf 2004}, {\em 598},~197 -- 202.
\newblock
  doi:{\changeurlcolor{black}\href{https://doi.org/https://doi.org/10.1016/j.physletb.2004.07.059}{\detokenize{https://doi.org/10.1016/j.physletb.2004.07.059}}}.

\bibitem[Gamermann and Oset(2009)]{PhysRevD.80.014003}
Gamermann, D.; Oset, E.
\newblock Isospin breaking effects in the $X(3872)$ resonance.
\newblock {\em Phys. Rev. D} {\bf 2009}, {\em 80},~014003.
\newblock
  doi:{\changeurlcolor{black}\href{https://doi.org/10.1103/PhysRevD.80.014003}{\detokenize{10.1103/PhysRevD.80.014003}}}.

\bibitem[Gamermann \em{et~al.}(2010)Gamermann, Nieves, Oset, and
  Arriola]{PhysRevD.81.014029}
Gamermann, D.; Nieves, J.; Oset, E.; Arriola, E.R.
\newblock Couplings in coupled channels versus wave functions: Application to
  the $X(3872)$ resonance.
\newblock {\em Phys. Rev. D} {\bf 2010}, {\em 81},~014029.
\newblock
  doi:{\changeurlcolor{black}\href{https://doi.org/10.1103/PhysRevD.81.014029}{\detokenize{10.1103/PhysRevD.81.014029}}}.

\bibitem[Ortega \em{et~al.}(2010)Ortega, Segovia, Entem, and
  Fern\'andez]{PhysRevD.81.054023}
Ortega, P.G.; Segovia, J.; Entem, D.R.; Fern\'andez, F.
\newblock Coupled channel approach to the structure of the $X(3872)$.
\newblock {\em Phys. Rev. D} {\bf 2010}, {\em 81},~054023.
\newblock
  doi:{\changeurlcolor{black}\href{https://doi.org/10.1103/PhysRevD.81.054023}{\detokenize{10.1103/PhysRevD.81.054023}}}.

\bibitem[Ortega \em{et~al.}(2013)Ortega, Entem, and
  Fern{\'{a}}ndez]{Ortega_2013}
Ortega, P.G.; Entem, D.R.; Fern{\'{a}}ndez, F.
\newblock Molecular structures in the charmonium spectrum: {theXYZpuzzle}.
\newblock {\em Journal of Physics G: Nuclear and Particle Physics} {\bf 2013},
  {\em 40},~065107.
\newblock
  doi:{\changeurlcolor{black}\href{https://doi.org/10.1088/0954-3899/40/6/065107}{\detokenize{10.1088/0954-3899/40/6/065107}}}.

\bibitem[Burns(2015)]{Burns2015}
Burns, T.J.
\newblock Phenomenology of Pc(4380)+, Pc(4450)+ and related states.
\newblock {\em The European Physical Journal A} {\bf 2015}, {\em 51},~152.
\newblock
  doi:{\changeurlcolor{black}\href{https://doi.org/10.1140/epja/i2015-15152-6}{\detokenize{10.1140/epja/i2015-15152-6}}}.

\bibitem[Guo \em{et~al.}(2019)Guo, Jing, Mei\ss{}ner, and
  Sakai]{PhysRevD.99.091501}
Guo, F.K.; Jing, H.J.; Mei\ss{}ner, U.G.; Sakai, S.
\newblock Isospin breaking decays as a diagnosis of the hadronic molecular
  structure of the ${P}_{c}(4457)$.
\newblock {\em Phys. Rev. D} {\bf 2019}, {\em 99},~091501.
\newblock
  doi:{\changeurlcolor{black}\href{https://doi.org/10.1103/PhysRevD.99.091501}{\detokenize{10.1103/PhysRevD.99.091501}}}.

\bibitem[Nieves and Pav\'on~Valderrama(2012)]{PhysRevD.86.056004}
Nieves, J.; Pav\'on~Valderrama, M.
\newblock Heavy quark spin symmetry partners of the $X(3872)$.
\newblock {\em Phys. Rev. D} {\bf 2012}, {\em 86},~056004.
\newblock
  doi:{\changeurlcolor{black}\href{https://doi.org/10.1103/PhysRevD.86.056004}{\detokenize{10.1103/PhysRevD.86.056004}}}.

\bibitem[Hidalgo-Duque \em{et~al.}(2013)Hidalgo-Duque, Nieves, and
  Valderrama]{PhysRevD.87.076006}
Hidalgo-Duque, C.; Nieves, J.; Valderrama, M.P.
\newblock Light flavor and heavy quark spin symmetry in heavy meson molecules.
\newblock {\em Phys. Rev. D} {\bf 2013}, {\em 87},~076006.
\newblock
  doi:{\changeurlcolor{black}\href{https://doi.org/10.1103/PhysRevD.87.076006}{\detokenize{10.1103/PhysRevD.87.076006}}}.

\bibitem[{Baru, V.} \em{et~al.}(2017){Baru, V.}, {Epelbaum, E.}, {Filin, A.
  A.}, {Hanhart, C.}, and {Nefediev, A.V.}]{Epel2017}
{Baru, V.}.; {Epelbaum, E.}.; {Filin, A. A.}.; {Hanhart, C.}.; {Nefediev,
  A.V.}.
\newblock Molecular partners of the X(3872) from heavy-quark spin symmetry: a
  fresh look.
\newblock {\em EPJ Web Conf.} {\bf 2017}, {\em 137},~06002.
\newblock
  doi:{\changeurlcolor{black}\href{https://doi.org/10.1051/epjconf/201713706002}{\detokenize{10.1051/epjconf/201713706002}}}.

\bibitem[Guo \em{et~al.}(2013)Guo, Hidalgo-Duque, Nieves, and
  Pav\'on~Valderrama]{PhysRevD.88.054007}
Guo, F.K.; Hidalgo-Duque, C.; Nieves, J.; Pav\'on~Valderrama, M.
\newblock Consequences of heavy-quark symmetries for hadronic molecules.
\newblock {\em Phys. Rev. D} {\bf 2013}, {\em 88},~054007.
\newblock
  doi:{\changeurlcolor{black}\href{https://doi.org/10.1103/PhysRevD.88.054007}{\detokenize{10.1103/PhysRevD.88.054007}}}.

\bibitem[Entem \em{et~al.}(2016)Entem, Ortega, and
  Fern\'andez]{doi:10.1063/1.4949442}
Entem, D.R.; Ortega, P.G.; Fern\'andez, F.
\newblock Partners of the X(3872) and heavy quark spin symmetry breaking.
\newblock {\em AIP Conference Proceedings} {\bf 2016}, {\em 1735},~060006.
\newblock
  doi:{\changeurlcolor{black}\href{https://doi.org/10.1063/1.4949442}{\detokenize{10.1063/1.4949442}}}.

\bibitem[Cincioglu \em{et~al.}(2016)Cincioglu, Nieves, Ozpineci, and
  Yilmazer]{Cincioglu2016}
Cincioglu, E.; Nieves, J.; Ozpineci, A.; Yilmazer, A.U.
\newblock Quarkonium Contribution to Meson Molecules.
\newblock {\em The European Physical Journal C} {\bf 2016}, {\em 76},~576.
\newblock
  doi:{\changeurlcolor{black}\href{https://doi.org/10.1140/epjc/s10052-016-4413-1}{\detokenize{10.1140/epjc/s10052-016-4413-1}}}.

\bibitem[Ortega \em{et~al.}(2018)Ortega, Segovia, Entem, and
  Fernández]{ORTEGA20181}
Ortega, P.G.; Segovia, J.; Entem, D.R.; Fernández, F.
\newblock Charmonium resonances in the 3.9 GeV/c2 energy region and the
  X(3915)/X(3930) puzzle.
\newblock {\em Physics Letters B} {\bf 2018}, {\em 778},~1 -- 5.
\newblock
  doi:{\changeurlcolor{black}\href{https://doi.org/https://doi.org/10.1016/j.physletb.2018.01.005}{\detokenize{https://doi.org/10.1016/j.physletb.2018.01.005}}}.

\bibitem[Batley \em{et~al.}(2006)Batley et~al.]{2006173}
Batley, J.; others.
\newblock Observation of a cusp-like structure in the $\pi^0\pi^0$ invariant
  mass distribution from $K^\pm \to \pi^\pm \pi^0 \pi^0$ decay and
  determination of the $\pi\pi$ scattering lengths.
\newblock {\em Physics Letters B} {\bf 2006}, {\em 633},~173 -- 182.
\newblock
  doi:{\changeurlcolor{black}\href{https://doi.org/https://doi.org/10.1016/j.physletb.2005.11.087}{\detokenize{https://doi.org/10.1016/j.physletb.2005.11.087}}}.

\bibitem[Budini and Fonda(1961)]{PhysRevLett.6.419}
Budini, P.; Fonda, L.
\newblock Pion-Pion Interaction from Threshold Anomalies in ${K}^{+}$ Decay.
\newblock {\em Phys. Rev. Lett.} {\bf 1961}, {\em 6},~419--421.
\newblock
  doi:{\changeurlcolor{black}\href{https://doi.org/10.1103/PhysRevLett.6.419}{\detokenize{10.1103/PhysRevLett.6.419}}}.

\bibitem[Cabibbo(2004)]{PhysRevLett.93.121801}
Cabibbo, N.
\newblock Determination of the $a_0-a_2$ Pion Scattering Length from
  ${K}^{+}\rightarrow\pi^{+}\pi^{0}\pi^{0}$ Decay.
\newblock {\em Phys. Rev. Lett.} {\bf 2004}, {\em 93},~121801.
\newblock
  doi:{\changeurlcolor{black}\href{https://doi.org/10.1103/PhysRevLett.93.121801}{\detokenize{10.1103/PhysRevLett.93.121801}}}.

\bibitem[Cabibbo and Isidori(2005)]{Cabibbo_2005}
Cabibbo, N.; Isidori, G.
\newblock Pion-pion scattering and the K$\rightarrow$3$\pi$ decay amplitudes.
\newblock {\em Journal of High Energy Physics} {\bf 2005}, {\em
  2005},~021--021.
\newblock
  doi:{\changeurlcolor{black}\href{https://doi.org/10.1088/1126-6708/2005/03/021}{\detokenize{10.1088/1126-6708/2005/03/021}}}.

\bibitem[Swanson(2015)]{PhysRevD.91.034009}
Swanson, E.S.
\newblock ${Z}_{b}$ and ${Z}_{c}$ exotic states as coupled channel cusps.
\newblock {\em Phys. Rev. D} {\bf 2015}, {\em 91},~034009.
\newblock
  doi:{\changeurlcolor{black}\href{https://doi.org/10.1103/PhysRevD.91.034009}{\detokenize{10.1103/PhysRevD.91.034009}}}.

\bibitem[Guo \em{et~al.}(2015)Guo, Hanhart, Wang, and Zhao]{PhysRevD.91.051504}
Guo, F.K.; Hanhart, C.; Wang, Q.; Zhao, Q.
\newblock Could the near-threshold $XYZ$ states be simply kinematic effects?
\newblock {\em Phys. Rev. D} {\bf 2015}, {\em 91},~051504.
\newblock
  doi:{\changeurlcolor{black}\href{https://doi.org/10.1103/PhysRevD.91.051504}{\detokenize{10.1103/PhysRevD.91.051504}}}.

\bibitem[Aaltonen \em{et~al.}(2011)Aaltonen et~al.]{Aaltonen:2011at}
Aaltonen, T.; others.
\newblock {Observation of the $Y(4140)$ structure in the $J/\psi\,\phi$ Mass
  Spectrum in $B^\pm\to J/\psi\,\phi K$ decays} {\bf 2011}.
\newblock  \href{http://xxx.lanl.gov/abs/1101.6058}{{\normalfont
  [arXiv:hep-ex/1101.6058]}}.

\bibitem[Abazov \em{et~al.}(2015)Abazov et~al.]{Abazov:2015sxa}
Abazov, V.M.; others.
\newblock {Inclusive Production of the X(4140) State in $p \overline p$
  Collisions at D0}.
\newblock {\em Phys. Rev. Lett.} {\bf 2015}, {\em 115},~232001,
  \href{http://xxx.lanl.gov/abs/1508.07846}{{\normalfont
  [arXiv:hep-ex/1508.07846]}}.
\newblock
  doi:{\changeurlcolor{black}\href{https://doi.org/10.1103/PhysRevLett.115.232001}{\detokenize{10.1103/PhysRevLett.115.232001}}}.

\bibitem[Chatrchyan \em{et~al.}(2014)Chatrchyan et~al.]{Chatrchyan:2013dma}
Chatrchyan, S.; others.
\newblock {Observation of a peaking structure in the $J/\psi \phi$ mass
  spectrum from $B^{\pm} \to J/\psi \phi K^{\pm}$ decays}.
\newblock {\em Phys. Lett.} {\bf 2014}, {\em B734},~261--281,
  \href{http://xxx.lanl.gov/abs/1309.6920}{{\normalfont
  [arXiv:hep-ex/1309.6920]}}.
\newblock
  doi:{\changeurlcolor{black}\href{https://doi.org/10.1016/j.physletb.2014.05.055}{\detokenize{10.1016/j.physletb.2014.05.055}}}.

\bibitem[Shen \em{et~al.}(2010)Shen et~al.]{Shen:2009vs}
Shen, C.P.; others.
\newblock {Evidence for a new resonance and search for the Y(4140) in the gamma
  gamma ---> phi J/psi process}.
\newblock {\em Phys. Rev. Lett.} {\bf 2010}, {\em 104},~112004,
  \href{http://xxx.lanl.gov/abs/0912.2383}{{\normalfont
  [arXiv:hep-ex/0912.2383]}}.
\newblock
  doi:{\changeurlcolor{black}\href{https://doi.org/10.1103/PhysRevLett.104.112004}{\detokenize{10.1103/PhysRevLett.104.112004}}}.

\bibitem[Lees \em{et~al.}(2015)Lees et~al.]{Lees:2014lra}
Lees, J.P.; others.
\newblock {Study of $B^{\pm,0} \to J/\psi K^+ K^- K^{\pm,0}$ and search for
  $B^0 \to J/\psi\phi$ at BABAR}.
\newblock {\em Phys. Rev.} {\bf 2015}, {\em D91},~012003,
  \href{http://xxx.lanl.gov/abs/1407.7244}{{\normalfont
  [arXiv:hep-ex/1407.7244]}}.
\newblock
  doi:{\changeurlcolor{black}\href{https://doi.org/10.1103/PhysRevD.91.012003}{\detokenize{10.1103/PhysRevD.91.012003}}}.

\bibitem[Aaij \em{et~al.}(2016)Aaij et~al.]{Aaij:2016iza}
Aaij, R.; others.
\newblock {Observation of $J/\psi\phi$ structures consistent with exotic states
  from amplitude analysis of $B^+\to J/\psi \phi K^+$ decays} {\bf 2016}.
\newblock  \href{http://xxx.lanl.gov/abs/1606.07895}{{\normalfont
  [arXiv:hep-ex/1606.07895]}}.

\bibitem[Ortega \em{et~al.}(2016)Ortega, Segovia, Entem, and
  Fern\'andez]{Ortega:2016hde}
Ortega, P.G.; Segovia, J.; Entem, D.R.; Fern\'andez, F.
\newblock {Canonical description of the new LHCb resonances}.
\newblock {\em Phys. Rev. D} {\bf 2016}, {\em 94},~114018,
  \href{http://xxx.lanl.gov/abs/1608.01325}{{\normalfont
  [arXiv:hep-ph/1608.01325]}}.
\newblock
  doi:{\changeurlcolor{black}\href{https://doi.org/10.1103/PhysRevD.94.114018}{\detokenize{10.1103/PhysRevD.94.114018}}}.

\bibitem[{Ortega, Pablo G.} \em{et~al.}(2019){Ortega, Pablo G.}, {Segovia,
  Jorge}, {Entem, David R.}, and {Fern\'andez, Francisco}]{refId0}
{Ortega, Pablo G.}.; {Segovia, Jorge}.; {Entem, David R.}.; {Fern\'andez,
  Francisco}.
\newblock The $Z_c$ structures in a coupled-channels model.
\newblock {\em Eur. Phys. J. C} {\bf 2019}, {\em 79},~78.
\newblock
  doi:{\changeurlcolor{black}\href{https://doi.org/10.1140/epjc/s10052-019-6552-7}{\detokenize{10.1140/epjc/s10052-019-6552-7}}}.

\bibitem[Ablikim \em{et~al.}(2013)Ablikim et~al.]{PhysRevLett.110.252001}
Ablikim, M.; others.
\newblock Observation of a Charged Charmoniumlike Structure in
  ${e}^{\mathbf{+}}{e}^{\mathbf{\ensuremath{-}}}\ensuremath{\rightarrow}{\ensuremath{\pi}}^{\mathbf{+}}{\ensuremath{\pi}}^{\mathbf{\ensuremath{-}}}J/\ensuremath{\psi}$
  at $\sqrt{s}\mathbf{=}4.26\text{ }\text{ }\mathrm{GeV}$.
\newblock {\em Phys. Rev. Lett.} {\bf 2013}, {\em 110},~252001.
\newblock
  doi:{\changeurlcolor{black}\href{https://doi.org/10.1103/PhysRevLett.110.252001}{\detokenize{10.1103/PhysRevLett.110.252001}}}.

\bibitem[Liu \em{et~al.}(2013)Liu et~al.]{PhysRevLett.110.252002}
Liu, Z.Q.; others.
\newblock Study of
  ${e}^{\mathbf{+}}{e}^{\mathbf{\ensuremath{-}}}\ensuremath{\rightarrow}{\ensuremath{\pi}}^{\mathbf{+}}{\ensuremath{\pi}}^{\mathbf{\ensuremath{-}}}J/\ensuremath{\psi}$
  and Observation of a Charged Charmoniumlike State at Belle.
\newblock {\em Phys. Rev. Lett.} {\bf 2013}, {\em 110},~252002.
\newblock
  doi:{\changeurlcolor{black}\href{https://doi.org/10.1103/PhysRevLett.110.252002}{\detokenize{10.1103/PhysRevLett.110.252002}}}.

\bibitem[Ablikim \em{et~al.}(2014)Ablikim et~al.]{PhysRevLett.112.022001}
Ablikim, M.; others.
\newblock Observation of a Charged
  ${(D{\overline{D}}^{*})}^{\ifmmode\pm\else\textpm\fi{}}$ Mass Peak in
  ${e}^{+}{e}^{\ensuremath{-}}\ensuremath{\rightarrow}\ensuremath{\pi}D{\overline{D}}^{*}$
  at $\sqrt{s}=4.26\text{ }\text{ }\mathrm{GeV}$.
\newblock {\em Phys. Rev. Lett.} {\bf 2014}, {\em 112},~022001.
\newblock
  doi:{\changeurlcolor{black}\href{https://doi.org/10.1103/PhysRevLett.112.022001}{\detokenize{10.1103/PhysRevLett.112.022001}}}.

\bibitem[Ablikim \em{et~al.}(2013)Ablikim et~al.]{PhysRevLett.111.242001}
Ablikim, M.; others.
\newblock Observation of a Charged Charmoniumlike Structure ${Z}_{c}(4020)$ and
  Search for the ${Z}_{c}(3900)$ in
  ${e}^{\mathbf{+}}{e}^{\mathbf{\ensuremath{-}}}\ensuremath{\rightarrow}{\ensuremath{\pi}}^{\mathbf{+}}{\ensuremath{\pi}}^{\mathbf{\ensuremath{-}}}{h}_{c}$.
\newblock {\em Phys. Rev. Lett.} {\bf 2013}, {\em 111},~242001.
\newblock
  doi:{\changeurlcolor{black}\href{https://doi.org/10.1103/PhysRevLett.111.242001}{\detokenize{10.1103/PhysRevLett.111.242001}}}.

\bibitem[Ablikim \em{et~al.}(2014)Ablikim et~al.]{PhysRevLett.113.212002}
Ablikim, M.; others.
\newblock Observation of
  ${e}^{+}{e}^{\ensuremath{-}}\ensuremath{\rightarrow}{\ensuremath{\pi}}^{0}{\ensuremath{\pi}}^{0}{h}_{c}$
  and a Neutral Charmoniumlike Structure ${Z}_{c}(4020{)}^{0}$.
\newblock {\em Phys. Rev. Lett.} {\bf 2014}, {\em 113},~212002.
\newblock
  doi:{\changeurlcolor{black}\href{https://doi.org/10.1103/PhysRevLett.113.212002}{\detokenize{10.1103/PhysRevLett.113.212002}}}.

\bibitem[Ablikim \em{et~al.}(2015)Ablikim et~al.]{Ablikim:2015swa}
Ablikim, M.; others.
\newblock {Confirmation of a charged charmoniumlike state $Z_c(3885)^{\mp}$ in
  $e^+e^-\to\pi^{\pm}(D\bar{D}^*)^\mp$ with double $D$ tag}.
\newblock {\em Phys. Rev.} {\bf 2015}, {\em D92},~092006,
  \href{http://xxx.lanl.gov/abs/1509.01398}{{\normalfont
  [arXiv:hep-ex/1509.01398]}}.
\newblock
  doi:{\changeurlcolor{black}\href{https://doi.org/10.1103/PhysRevD.92.092006}{\detokenize{10.1103/PhysRevD.92.092006}}}.

\bibitem[Ablikim \em{et~al.}(2017)Ablikim et~al.]{Collaboration:2017njt}
Ablikim, M.; others.
\newblock {Determination of the Spin and Parity of the $Z_c(3900)$}.
\newblock {\em Phys. Rev. Lett.} {\bf 2017}, {\em 119},~072001,
  \href{http://xxx.lanl.gov/abs/1706.04100}{{\normalfont
  [arXiv:hep-ex/1706.04100]}}.
\newblock
  doi:{\changeurlcolor{black}\href{https://doi.org/10.1103/PhysRevLett.119.072001}{\detokenize{10.1103/PhysRevLett.119.072001}}}.

\bibitem[Ablikim \em{et~al.}(2014)Ablikim et~al.]{Ablikim:2013emm}
Ablikim, M.; others.
\newblock {Observation of a charged charmoniumlike structure in $e^+e^- \to
  (D^{*} \bar{D}^{*})^{\pm} \pi^\mp$ at $\sqrt{s}=4.26$GeV}.
\newblock {\em Phys. Rev. Lett.} {\bf 2014}, {\em 112},~132001,
  \href{http://xxx.lanl.gov/abs/1308.2760}{{\normalfont
  [arXiv:hep-ex/1308.2760]}}.
\newblock
  doi:{\changeurlcolor{black}\href{https://doi.org/10.1103/PhysRevLett.112.132001}{\detokenize{10.1103/PhysRevLett.112.132001}}}.

\end{thebibliography}

\end{document}